\documentclass
[preprint,superscriptaddress,showpacs,showkeys,prb,eqsecnum]{revtex4}%
\usepackage{amsfonts}
\usepackage{amsmath}
\usepackage{amssymb}
\usepackage{graphicx}%
\providecommand{\U}[1]{\protect\rule{.1in}{.1in}}
\newlength{\defaultparindent}
\newenvironment{annotation Text}{}{}
\begin{document}
\title{Perturbative Theory of Grazing-Incidence Diffuse Nuclear Resonant Scattering
of Synchrotron Radiation}
\author{L. De\'{a}k}
\email{deak@rmki.kfki.hu}
\affiliation{KFKI Research Institute for Particle and Nuclear Physics, P.O.B. 49, H-1525
Budapest, Hungary}
\author{L. Botty\'{a}n}
\affiliation{KFKI Research Institute for Particle and Nuclear Physics, P.O.B. 49, H-1525
Budapest, Hungary}
\author{D.L. Nagy}
\affiliation{KFKI Research Institute for Particle and Nuclear Physics, P.O.B. 49, H-1525
Budapest, Hungary}
\author{H. Spiering}
\affiliation{Institut f\"{u}r Anorganische und Analytische Chemie, Johannes Gutenberg
Universit\"{a}t Mainz, Staudinger Weg 9, D-55099 Mainz, Germany}
\author{Yu.N. Khaidukov}
\affiliation{Frank Laboratory of Neutron Physics, Joint Institute for Nuclear
Research,141980, Dubna, Moscow Region, Russia}
\author{Y.~Yoda }
\affiliation{SPring-8 JASRI, 1-1-1 Kouto Mikazuki-cho Sayo-gun Hyogo 679-5198, Japan}
\date{\today}

\begin{abstract}
Theoretical description of off-specular grazing-incidence nuclear resonant
scattering of synchrotron radiation (Synchrotron M\"{o}ssbauer Reflectometry,
SMR) is presented. The recently developed SMR, similarly to polarized neutron
reflectometry (PNR), is an analytical tool for the determination of isotopic
and magnetic structure of thin films and multilayers. It combines the
sensitivity of M\"{o}ssbauer spectroscopy to hyperfine interactions and the
depth selectivity of x-ray reflectometry. Specular reflection provides
information on the depth profile, while off-specular scattering on the lateral
structure of scattering layers. Off-specular SMR and PNR intensity formulae of
a rather general multilayer with different domains, based on a Distorted
Incident-Wave Approximation (DIWA) are presented. The Distorted-Wave Born
Approximation (DWBA) results are given in an Appendix. Physical and numerical
implications, why using DIWA, are explained. The temporal character of SMR
imposes specific differences between SMR and PNR. In order to reveal the
limits of DIWA and to compare the two analytical methods, two-dimensional
diffuse SMR and PNR maps of an antiferromagnetic multilayer are calculated and
critically compared. Experimental '$\omega-2\theta$' SMR map of a periodic
$\left[  \mathrm{Fe}/\mathrm{Cr}\right]  _{20}$ multilayer is presented and
compared with simulations by the present theory.

\end{abstract}
\keywords{Reflectometry, grazing incidence, diffuse scattering, distorted wave
approximation, nuclear resonance scattering, neutron scattering, multilayers }
\pacs{PACS: 42.25-p, 61.10.Kw, 61.12.Ha, 75.25.+z}
\preprint{HEP/123-qed}
\maketitle


\section{Introduction}

\bigskip Grazing-incidence reflection from flat surfaces of
x-rays,\cite{Stoev1999,Lax51,Zhou95,Daillant}
neutrons\cite{Lax51,Zhou95,Daillant,Felcher93,Majkrzak91,Sinha91} and of soft
x-rays near the absorption edges\cite{Daillant,Hannon88,Whan94} (x-ray,
neutron and resonant magnetic x-ray reflectometry, respectively) have been
widely used to investigate the chemical, isotopic and magnetic structure of
thin films and multilayers (ML).\cite{Hase00,Lauter00,Langridge00} Making use
of the sensitivity of nuclear resonant scattering (NRS) of synchrotron
radiation (SR) to hyperfine interactions, another analytical method has been
developed that combines the sensitivity of M\"{o}ssbauer spectroscopy to
hyperfine interactions with the depth information yielded by reflectometry.
NRS of SR is performed in the time domain. All nuclear resonances (of
typically a few neV linewidth) are excited simultaneously by a SR pulse of a
few meV bandwidth. Time-differential NRS of SR contains the hyperfine
interaction information in the quantum-beat pattern of the time response that
follows the excitation of the system by the synchrotron pulse. Counting,
however, all delayed (i.e., nuclear resonant scattered) photons as a function
of the grazing angle of incidence is a special kind of x-ray reflectometry
that we shall call henceforth Synchrotron M\"{o}ssbauer Reflectometry
(SMR).\cite{Nagy99,Deak02} This method is similar to polarized neutron
reflectometry (PNR) and yields integral hyperfine depth profile and
superstructure information. Specular SMR has by now become an established
technique.\cite{Toellner95,Chumakov99,Deak02,Rohlsberger03}

The specularly reflected radiation from a stratified system is insensitive to
the lateral structure; it depends solely on the lateral averages of the
material parameters\cite{Lax51} for the coherence volume.\cite{Mandel-Wolf} If
one uses radiation of infinite coherence length, the contrast of laterally
compensated domains (e.g. the magnetic contrast of an antiferromagnetic ML
stack) would be absent from the specular reflectivity. However, this is not
the case, since the coherent averaging has to be performed for the finite
coherence volume\cite{Mandel-Wolf} of the radiation determined by the
experimental setup and such contributions are to be added
incoherently.\cite{Toperverg01a}

For studying lateral inhomogeneities, such as structural roughness, magnetic
domains, etc., diffuse scattering, i.e. off-specular reflectometry is used.
('Diffuse' and 'off-specular' will be used as synonyms throughout this present
paper.) Off-specular non-polarized\cite{Langridge00} and polarized neutron
reflectometry,\cite{Felcher93,Lauter00} soft-x-ray resonant magnetic diffuse
scattering \cite{Hase00} and, recently, off-specular SMR\cite{Nagy02a} have
been used to estimate the domain-size distribution and to follow domain
transformations in antiferromagnetically (AF)-coupled magnetic MLs.

The optical theory of \textit{specular} SMR has been published using several,
somewhat different approaches,
\cite{Rohlsberger03,Deak01,Afanasjev65,Hannon69,Hannon85,Irkajev93,Irkajev93b,Deak96,Rohlsberger99a,Rohlsberger99b}
however, to our knowledge, the present attempt is the first one to describe
nuclear resonant diffuse scattering. Distorted-Wave Born
Approximation\cite{Sinha88} (DWBA) was used earlier by several
authors\cite{Toperverg01a,Toperverg01b,Lauter94,Ruhm99} to describe
off-specular scattering of \textit{neutrons} by a rough surface. DWBA
perturbatively determines the (off-specular)
field\cite{Vineyard82,Dosch,Ljungsdahl,Zhou95} around the specular field, the
latter being easily calculated, even for general stratified media, by suitable
matrix
methods.\cite{Toperverg01a,Toperverg01b,Felcher87,Paratt54,Majkrzak89,Rohlsberger03,Deak01}%

The choice of DWBA to elaborate \textit{off-specular} SMR data seems,
therefore, obvious. However, unlike PNR, in the SMR case the scattering
potential is strongly energy dependent. Moreover, SMR is detected in the time
domain. Therefore the existing theory can not be directly applied. Since the
2D diffuse maps have to be calculated for typically a thousand time- (or
energy) channels, therefore, in practice, the anyway long calculation time of
the diffuse maps by an orthodox DWBA would multiply by a thousand and become
intolerable long for SMR.

The aim of the present paper is to present an alternative general evaluation
algorithm for off-specular SMR. A distorted--wave approximation will be used,
which is --- except for \textit{exit }angles in the total reflection region
--- accurate enough and, at the same time, it is capable of handling the
immense computational problem in a reasonable period of time. For reasons
explained in Sec. \ref{off-specular-SMR}, we shall call this approximation
'Distorted Incident-Wave Approximation' (DIWA).

Starting from Lax' general theory\cite{Lax51} and from the common optical
formalism of polarized neutron and M\"{o}ssbauer reflectometry,\cite{Deak01}
an expression will be deduced for diffuse scattering of electromagnetic and/or
quantum mechanical particle waves on laterally inhomogeneous stratified media.
From the point of view of specular reflection, the set of discrete atomic
scattering centers will be replaced by a homogeneous index of refraction $n$
and the scattering problem will be traced back to solving the wave equation
(far from Bragg reflections).\cite{Lax51} In the rest of the paper the
grazing-incidence limit is studied, for which the index of refraction
approximation is
justified.\cite{Rohlsberger03,Hannon85,Deak96,Rohlsberger99a,Rohlsberger99b}

The material of the paper is organized as follows. In Sec.
\ref{off-specular-scattering},\ we present a distorted--wave approximation
without treating the energy dependence of the problem. The presented method is
a generalization of Vineyard's approximation\cite{Vineyard82} for the case of
MLs having domain-like inhomogeneities with absorptive anisotropic index of
refractions (scattering potentials). In Sec. \ref{off-specular-SMR}
time-differential and time-integral diffuse intensities are derived for quite
general inter- and intra-layer correlations using a domain correlation
matrix.\emph{\ }The coherent-field solution and the cumulative transmissivity
for an arbitrary depth is calculated in Appendices \ref{Appendix1} and
\ref{Appendix2}, respectively. The DWBA formulae for the delayed reflectivity
are given in Appendix \ref{Appendix3}. In Sec. \ref{Applications} we provide
numerical simulations in the DIWA approach for both PNR\ and SMR. Specular and
diffuse PNR and SMR scans and maps are compared and the special features of
the methods are discussed. In Sec. \ref{Experimental results and discussion}
the developed DIWA\ SMR theory is applied to extract the average
antiferromagnetic domain size from experimental SMR $\omega-2\theta$ maps of
an [Fe/Cr] epitaxial ML.

\section{Off-specular scattering\label{off-specular-scattering}}

The model systems of the present paper are MLs and thin films, having lateral
inhomogeneities on the mesoscopic scale. Unlike the case of surface roughness,
these inhomogeneities (domains) will be assumed to be much larger than the
atomic distances. In each homogeneous part around position $\mathbf{r}$ an
index of refraction $n\left(  \mathbf{r}\right)  $ is defined. Since the
elements of $n$ for both slow neutrons and X-rays differ only slightly
(typically $10^{-2}-10^{-5}$) from that of the $2\times2$ unit matrix $I$, the
small parameter, namely the generalized susceptibility, $\chi\left(
\mathbf{r}\right)  =2\left[  n\left(  \mathbf{r}\right)  -I\right]  $ can be
conveniently defined.\cite{Deak01,Deak96,Nickel2001}

Using the index-of-refraction approximation, in each homogeneous part of the
system, the solution of the homogeneous wave equation%
\begin{equation}
\left[  \Delta+k^{2}I\right]  \Psi\left(  \mathbf{r}\right)  =-k^{2}%
\chi\left(  \mathbf{r}\right)  \Psi\left(  \mathbf{r}\right)  \label{1}%
\end{equation}
yields $\Psi\left(  \mathbf{r}\right)  $, representing the two components of
the photon field or the neutron quantum mechanical spinor state at position
$\mathbf{r}$\textbf{,} with $k$ being the wave number in vacuum. $\chi$ is
simply related to the coherent forward-scattering amplitude $f$ by $\chi
=\frac{4\pi N}{k^{2}}f$ where $N$ is the number of scattering centers per unit
volume.\cite{Deak01,Deak96} For photons $f=f_{\mathrm{e}}+f_{\mathrm{n}} $ is
the sum of the electronic and nuclear scattering amplitudes.\cite{Hannon85b}
For neutrons $f=f_{\mathrm{nuc}}+f_{\mathrm{magn}}$ is the sum of the nuclear
and magnetic scattering lengths. In Eq. (\ref{1}) $k^{2}\chi$ plays the role
of the optical scattering potential (scattering-length density in the neutron
literature). For photons $\chi$ is the susceptibility and, for sake of
simplicity, throughout the paper we shall use this term in its general sense.

Describing the system as a stack of (possibly laterally inhomogeneous) layers,
we compose the susceptibility by%

\begin{equation}
\chi\left(  \mathbf{r}\right)  =\sum_{l=1}^{S}\chi_{l}\left(  \mathbf{r}%
_{\parallel}\right)  , \label{2}%
\end{equation}
as the sum of the susceptibility functions of the individual layers $l$ ($l$
$=1$..$S$, the last layer $S$ being the substrate) depending solely on the
in-plane coordinate $\mathbf{r}_{\parallel}$.

If the homogeneous parts of the system are large compared to the wavelength we
may assume that the exact solution $\Psi\left(  \mathbf{r}\right)  $ is close
to the solution $\Psi_{\mathrm{coh}}\left(  \mathbf{r}\right)  $ of the
coherent (specular) field equation\cite{Lax51}
\begin{equation}
\left[  \Delta+k^{2}I\right]  \Psi_{\mathrm{coh}}\left(  \mathbf{r}\right)
=-k^{2}\sum_{l=1}^{S}\overline{\chi}_{l}\Psi_{\mathrm{coh}}\left(
\mathbf{r}\right)  \text{,} \label{3}%
\end{equation}
which is obtained from Eq. (\ref{1}) by replacing the susceptibilities
$\chi_{l}\left(  \mathbf{r}_{\parallel}\right)  $ by the average
susceptibility $\overline{\chi}_{l}$ of each layer $l$. In order to deduce a
perturbative equation, the sum $-k^{2}\sum\overline{\chi}_{l}\Psi\left(
\mathbf{r}\right)  $ is added and subtracted on the right-hand side of Eq.
(\ref{1})
\begin{equation}
\left[  \Delta+k^{2}I\right]  \Psi\left(  \mathbf{r}\right)  =-k^{2}\sum
_{l=1}^{S}\overline{\chi}_{l}\Psi\left(  \mathbf{r}\right)  -k^{2}\sum
_{l=1}^{S}\left[  \chi_{l}\left(  \mathbf{r}_{\parallel}\right)
-\overline{\chi}_{l}\right]  \Psi\left(  \mathbf{r}\right)  \text{.} \label{4}%
\end{equation}
For homogeneous layers $\chi_{l}\left(  \mathbf{r}_{\parallel}\right)
=\overline{\chi}_{l}$ , \textit{i.e.}, the second sum vanishes on the right,
so that Eq. (\ref{4}) reduces to Eq. (\ref{3}), the basic equation of specular
reflectometry.\cite{Ruhm99,Deak01,Deak96,Spiering00}

The general solutions of Eq. (\ref{4}) are searched for in a form%
\begin{equation}
\Psi\left(  \mathbf{r}\right)  =\Psi_{\mathrm{coh}}\left(  \mathbf{r}\right)
+\Psi_{\mathrm{off}}\left(  \mathbf{r}\right)  \label{5}%
\end{equation}
where $\Psi_{\mathrm{coh}}\left(  \mathbf{r}\right)  $ is the coherent field,
which vanishes in any non-specular direction, and $\Psi_{\mathrm{off}}\left(
\mathbf{r}\right)  $ is the off-specular field. Substituting Eq. (\ref{5})
into Eq. (\ref{4}) and taking into account Eq. (\ref{3}) we obtain
\begin{equation}
\left[  \Delta+k^{2}I\right]  \Psi_{\mathrm{off}}\left(  \mathbf{r}\right)
=-k^{2}\sum_{l=1}^{S}\left[  \chi_{l}\left(  \mathbf{r}_{\parallel}\right)
-\overline{\chi}_{l}\right]  \Psi_{\mathrm{coh}}\left(  \mathbf{r}\right)
-k^{2}\sum_{l=1}^{S}\chi_{l}\left(  \mathbf{r}_{\parallel}\right)
\Psi_{\mathrm{off}}\left(  \mathbf{r}\right)  \text{.} \label{6}%
\end{equation}
The coherent field $\Psi_{\mathrm{coh}}\left(  \mathbf{r}\right)  $, solution
of Eq. (\ref{3}), is obtained by the optical method \cite{Deak01} as%
\begin{equation}
\Psi_{\mathrm{coh}}\left(  \mathbf{k},\mathbf{r}\right)  =T\left(  k_{\perp
},r_{\perp}\right)  \Psi^{\mathrm{in}}\mathrm{\exp}\left(  i\mathbf{k}%
_{\parallel}\mathbf{r}_{\parallel}\right)  \label{7}%
\end{equation}
(see Appendix \ref{Appendix1}) where $\bot$ and $\parallel$ denote the
plane-perpendicular and in-plane components of the respective vectors and
$\Psi^{\mathrm{in}}$ is the amplitude of the incident plane wave of wave
vector $\mathbf{k}=\left(  k_{\perp},\mathbf{k}_{\parallel}\right)  $. Here we
introduced the 'cumulative transmittance'\ of the reflecting film from surface
to a depth $r_{\perp}$ by:%
\begin{equation}
T\left(  k_{\perp},r_{\perp}\right)  =L^{\left[  21\right]  }\left(  k_{\perp
},r_{\perp}\right)  \left[  I-R_{\mathrm{sp}}\left(  k_{\perp}\right)
\right]  +L^{\left[  22\right]  }\left(  k_{\perp},r_{\perp}\right)  \left[
I+R_{\mathrm{sp}}\left(  k_{\perp}\right)  \right]  . \label{8}%
\end{equation}
$R_{\mathrm{sp}}\left(  k_{\perp}\right)  $ is the $2\times2$ specular
reflectivity matrix of the system \cite{Deak01}, $L^{\left[  21\right]
}\left(  k_{\perp},r_{\perp}\right)  $ and $L^{\left[  22\right]  }\left(
k_{\perp},r_{\perp}\right)  $ are the respective $2\times2$ submatrices of the
$4\times4$ characteristic matrix,\cite{Deak01,Deak96} $L$ at depth $r_{\perp}$
for an incoming plane wave defined by $\mathbf{k}$.

The physical interpretation of the inhomogeneous wave equation (\ref{6}) is
seen from its right-hand side where the first term gives the source of the
off-specular radiation and the second term describes the off-specular field
scattered by the entire ML. The off-specular field arises from the coherent
field at the lateral inhomogeneities, \textit{i.e.} from the regions where the
susceptibility $\chi$ differs from its average value $\overline{\chi}$. Eq.
(\ref{6}) can be solved iteratively to the prescribed accuracy.

As a first approximation, the second term of the right-hand side of Eq.
(\ref{6}) is neglected so that the solution can be obtained using the
Green-function technique,
\begin{equation}
\Psi_{\mathrm{off}}\left(  \mathbf{k},\mathbf{r}\right)  =\frac{k^{2}}{4\pi
}\sum_{l}%
{\displaystyle\int}
\mathrm{d}^{3}\mathbf{r}^{\prime}\,\frac{\exp\left(  ikR\right)  }{R}\left[
\chi_{l}\left(  \mathbf{r}_{\parallel}^{\prime}\right)  -\overline{\chi_{l}%
}\right]  \Psi_{\mathrm{coh}}\left(  \mathbf{k},\mathbf{r}^{\prime}\right)
\text{,} \label{9}%
\end{equation}
where $R=\left\vert \mathbf{r}-\mathbf{r}^{\prime}\right\vert .$ The
approximation requires $\left\Vert \Psi_{\mathrm{coh}}\left(  \mathbf{r}%
\right)  \right\Vert \gg\left\Vert \Psi_{\mathrm{off}}\left(  \mathbf{r}%
\right)  \right\Vert $, which condition is fulfilled for magnetic MLs of large
enough homogeneous domain size in the vicinity of the specular
direction\cite{Nagy02a,Lauter94} so that the exact solution $\Psi\left(
\mathbf{r}\right)  $ is close to the coherent field $\Psi_{\mathrm{coh}%
}\left(  \mathbf{r}\right)  $. When neglecting the second term in Eq.
(\ref{6}) the scattering of the off-specular field is neglected. Therefore the
present distorted-wave approximation breaks down for \textit{exit angles} near
the critical angle of total reflection.

Far from the scatterer, the Fraunhofer approximation%
\begin{equation}
\frac{\exp\left(  ikR\right)  }{R}\approx\frac{\exp\left(  ikr\right)  }%
{r}\exp\left(  -i\mathbf{k^{\prime}r}^{\prime}\right)  \label{10}%
\end{equation}
is applied with $\mathbf{k}^{\prime}$ being the wave number vector of the
emerging plane wave, with which the final expression for the off-specular
field is
\begin{equation}
\Psi_{\mathrm{off}}\left(  \mathbf{k},\mathbf{r}=\frac{\mathbf{k}^{\prime}}%
{k}r\right)  =\sqrt{\frac{\pi}{2}}\frac{k^{2}}{r}\exp\left(  ikr\right)
\sum_{l}S_{l}\left(  \mathbf{K}_{\parallel}\right)  T_{l}\left(  k_{\perp
},k_{\perp}^{\prime}\right)  \Psi^{\mathrm{in}}\text{,} \label{11}%
\end{equation}
with $\mathbf{K}_{\parallel}$ being the in-plane component of the momentum
transfer vector $\mathbf{K}=\mathbf{k}^{\prime}-\mathbf{k}$, and%
\begin{equation}
T_{l}\left(  k_{\perp},k_{\perp}^{\prime}\right)  =\frac{1}{\sqrt{2\pi}}%
\int\limits_{Z_{l}}\mathrm{d}r_{\perp}\exp\left(  -ik_{\perp}^{\prime}%
r_{\perp}\right)  T\left(  k_{\perp},r_{\perp}\right)  \label{12}%
\end{equation}
is the Fourier integral over the one-dimensional interval $Z_{l}$ of layer $l
$. The expression
\begin{equation}
S_{l}\left(  \mathbf{K}_{\parallel}\right)  =\frac{1}{2\pi}\int\mathrm{d}%
^{2}\mathbf{r}_{\parallel}\,\exp\left(  -i\mathbf{K}_{\parallel}%
\,\mathbf{r}_{\parallel}\right)  \left[  \chi_{l}\left(  \mathbf{r}%
_{\parallel}\right)  -\overline{\chi_{l}}\right]  \label{13}%
\end{equation}
is the two-dimensional Fourier transform of $\chi_{l}\left(  \mathbf{r}%
_{\parallel}\right)  -\overline{\chi}$. $T_{l}\left(  k_{\perp},k_{\perp
}^{\prime}\right)  $ that can be analytically calculated (See Appendix
\ref{Appendix2}).

The off-specular scattered intensity $I_{\mathrm{off}}=\left(  \Psi
_{\mathrm{off}},\Psi_{\mathrm{off}}\right)  $ is
\begin{equation}
I_{\mathrm{off}}=\frac{\pi k^{4}}{2r^{2}}\sum_{ll^{\prime}}\left(
\Psi^{\mathrm{in}},T_{l}^{\dagger}\left(  k_{\perp},k_{\perp}^{\prime}\right)
S_{l}^{\dagger}\left(  \mathbf{K}_{\parallel}\right)  S_{l^{\prime}}\left(
\mathbf{K}_{\parallel}\right)  T_{l^{\prime}}\left(  k_{\perp},k_{\perp
}^{\prime}\right)  \Psi^{\mathrm{in}}\right)  \text{,} \label{14}%
\end{equation}
which, for arbitrary incident polarization, can be rewritten as
\begin{equation}
I_{\mathrm{off}}=\frac{\pi k^{4}}{2r^{2}}\sum_{ll^{\prime}}\operatorname{Tr}%
\left[  T_{l}^{\dagger}\left(  k_{\perp},k_{\perp}^{\prime}\right)
S_{l}^{\dagger}\left(  \mathbf{K}_{\parallel}\right)  S_{l^{\prime}}\left(
\mathbf{K}_{\parallel}\right)  T_{l^{\prime}}\left(  k_{\perp},k_{\perp
}^{\prime}\right)  \,\rho\right]  , \label{15}%
\end{equation}
where $\rho$ is the polarization density matrix of the incident
radiation.\cite{Blume68} From the convolution theorem, it follows that the
Fourier transform ${\mathfrak{C}}_{ll^{\prime}}\left(  \mathbf{R}_{\parallel
}\right)  $ of
\begin{equation}
C_{ll^{\prime}}\left(  \mathbf{K}_{\parallel}\right)  =\left(  2\pi\right)
S_{l}^{\dagger}\left(  \mathbf{K}_{\parallel}\right)  S_{l^{\prime}}\left(
\mathbf{K}_{\parallel}\right)  \label{16}%
\end{equation}
is the cross-correlation function of the susceptibilities between layers $l$
and $l^{\prime}$
\begin{equation}
{\mathfrak{C}}_{ll^{\prime}}\left(  \mathbf{R}_{\parallel}\right)
=\int\mathrm{d}^{2}\mathbf{r}_{\parallel}\,\left[  \chi_{l}\left(
\mathbf{R}_{\parallel}+\mathbf{r}_{\parallel}\right)  -\overline{\chi}%
_{l}\right]  ^{\dagger}\left[  \chi_{l^{\prime}}\left(  \mathbf{r}_{\parallel
}\right)  -\overline{\chi}_{l^{\prime}}\right]  . \label{17}%
\end{equation}
The final result then becomes
\begin{equation}
I_{\mathrm{off}}=\frac{k^{4}}{4r^{2}}\sum_{ll^{\prime}}\operatorname{Tr}%
\left[  T_{l}^{\dagger}\left(  k_{\perp},k_{\perp}^{\prime}\right)
C_{ll^{\prime}}\left(  \mathbf{K}_{\parallel}\right)  T_{l^{\prime}}\left(
k_{\perp},k_{\perp}^{\prime}\right)  \,\rho\right]  \text{,} \label{18}%
\end{equation}
a convenient expression for randomly distributed lateral inhomogeneities. We
note that the off-specular intensity $I_{\mathrm{off}}=I_{\mathrm{off}}\left(
\mathbf{K}_{\parallel},k_{\perp},k_{\perp}^{\prime}\right)  $ is a function of
$\mathbf{K}_{\parallel}$, $k_{\perp}$ and $k_{\perp}^{\prime}$, a notation
dropped in the calculations. The corresponding values of $\mathbf{K}%
_{\parallel}$, $k_{\perp}$ and $k_{\perp}^{\prime}$ can be given for the
chosen experimental geometry. The diffuse intensity expression in Eq.
(\ref{18}) differs from that of DWBA,\cite{Sinha88} \emph{\ }an issue to be
discussed in Sec. \ref{off-specular-SMR} and in Appendix \ref{Appendix3}.

A possible experimental realization of the off-specular reflectometry is the
so-called '$\omega$-scan' geometry where the detector position is set to
$2\theta$ and the sample orientation $\omega$ on the goniometer is varied with
the sample normal remaining in the scattering plane. For $\omega$-scans, the
in-plane components of the momentum transfer vector $K_{\parallel}$ and the
plane-perpendicular component of the wave vector of the emerging wave
$k_{\perp}^{\prime}$can be expressed by the grazing angle $\theta$ and by
$\omega$:%
\begin{equation}
K_{\parallel}=2k\sin\theta\sin\left(  \omega-\theta\right)  , \label{31}%
\end{equation}%
\begin{equation}
k_{\perp}=k\sin\omega, \label{31pl}%
\end{equation}%
\begin{equation}
k_{\perp}^{\prime}=-k\sin\left(  2\theta-\omega\right)  . \label{32}%
\end{equation}
One may observe that $K_{\parallel}$ vanishes at the specular condition
$\omega$ $=\theta$. A two-dimensional representation of the full diffuse
scatter is the '$\omega-2\theta$'-map. A further,\emph{\ }widely used,
arrangement is the so-called 'detector scan' geometry in which the angle of
incidence $\theta_{\mathrm{in}}$ is fixed and the scattered intensity is
recorded as a function of the exit angle $\theta_{\mathrm{out}}$. In case of a
detector scan, Eqs. (\ref{31}) and (\ref{34}) read%
\begin{equation}
K_{\parallel}=k\left(  \cos\theta_{\mathrm{out}}-\cos\theta_{\mathrm{in}%
}\right)  , \label{33}%
\end{equation}%
\begin{equation}
k_{\perp}=k\sin\theta_{\mathrm{in}}\text{,} \label{33pl}%
\end{equation}%
\begin{equation}
k_{\perp}^{\prime}=-k\sin\theta_{\mathrm{out}}\text{,} \label{34}%
\end{equation}
with the specular condition being $\theta_{\mathrm{out}}=\theta_{\mathrm{in}}%
$. An alternative representation of the full diffuse scatter is the
'$\theta_{\mathrm{in}}-$ $\theta_{\mathrm{out}}$' map.

\section{Off-specular SMR\label{off-specular-SMR}}

\subsection{Time-differential scattered intensity}

Eq. (\ref{18}) for the off-specular intensity is equally valid for neutron and
x-ray scattering where the time of the scattering process is negligible.
However, in case of nuclear resonant scattering we detect the time response
\begin{equation}
\Psi_{\mathrm{off}}\left(  \mathbf{r,}t\right)  =\frac{1}{\hbar\sqrt{2\pi}%
}\int\limits_{-\infty}^{\infty}\mathrm{d}E\,\Psi_{\mathrm{off}}\left(
\mathbf{r,}E\right)  \exp\left(  -iEt/\hbar\right)  \label{19}%
\end{equation}
after the synchrotron pulse,\cite{Trammell79} which is the Fourier transform
of the energy-dependent off-specular field. Close to a M\"{o}ssbauer
resonance, both the susceptibilities $\chi_{l}\left(  \mathbf{r}_{\parallel
}^{\prime},E\right)  -\overline{\chi_{l}}\left(  E\right)  $ and the coherent
field $\Psi_{\mathrm{coh}}\left(  \mathbf{r}^{\prime},E\right)  $ are strongly
energy-dependent.\cite{Trammell79} Therefore, through Eqs. (\ref{8}),
(\ref{12}) and (\ref{13}), the $S_{l}\left(  \mathbf{K}_{\parallel},E\right)
$ and the $T_{l}\left(  k_{\perp},k_{\perp}^{\prime},E\right)  $ quantities
carry an energy dependence, too. Consequently, Eq. (\ref{18}) can no longer be
applied to calculate the off-specular intensity.

A possible workaround of this problem is to define a distribution function
$\Omega_{l}^{\mu}\left(  \mathbf{r}_{\parallel}\right)  $ of homogeneous
regions of type $\mu=1,..,M$ of layer $l.$ This function $\Omega_{l}^{\mu
}\left(  \mathbf{r}_{\parallel}\right)  $ characterizes the homogeneous
regions of layer $l$ of an energy-dependent susceptibility $\chi^{\mu}\left(
E\right)  $. Over the region of type $\mu$ in layer $l$ the distribution
function $\Omega_{l}^{\mu}\left(  \mathbf{r}_{\parallel}\right)  =1$ otherwise
$\Omega_{l}^{\mu}\left(  \mathbf{r}_{\parallel}\right)  =0$. Any point
$\mathbf{r}_{\parallel}$\ along the surface of layer $l$\ is related to one
domain type, therefore%
\begin{equation}
\sum_{\mu}\Omega_{l}^{\mu}\left(  \mathbf{r}_{\parallel}\right)  =1.
\label{19p}%
\end{equation}
The total inhomogeneous susceptibility is the sum
\begin{equation}
\chi\left(  \mathbf{r,}E\right)  =\sum_{\mu=1}^{M}\sum_{l=1}^{S}\Omega
_{l}^{\mu}\left(  \mathbf{r}_{\parallel}\right)  \chi^{\mu}\left(  E\right)
\label{20}%
\end{equation}
where the space- and energy-dependent terms in $\chi\left(  \mathbf{r,}%
E\right)  $ have been separated. The average susceptibility within layer $l$
is
\begin{equation}
\overline{\chi_{l}}\left(  E\right)  =\sum_{\mu=1}^{M}\eta_{l}^{\mu}\chi^{\mu
}\left(  E\right)  \label{21}%
\end{equation}
where the fractional domain area is\emph{\ }%
\begin{equation}
\eta_{l}^{\mu}=A_{l}^{\mu}/A \label{defeta}%
\end{equation}
\textit{i.e.}, the ratio of $A_{l}^{\mu}=\int d^{2}\mathbf{r}_{\parallel
}\,\Omega_{l}^{\mu}\left(  \mathbf{r}_{\parallel}\right)  $, the total area of
the homogeneous part of type $\mu$\ within layer $l$, and $A,$\ the area of
the ML. Since the domains fully cover the layers, the condition%
\begin{equation}
\sum_{\mu}\eta_{l}^{\mu}=1 \label{21a}%
\end{equation}
is fulfilled. Using (\ref{20}) and (\ref{21}), Eq. (\ref{13}) becomes
\begin{equation}
S_{l}\left(  \mathbf{K}_{\parallel},E\right)  \mathbf{=}\sum_{\mu=1}^{M}%
W_{l}^{\mu}\left(  \mathbf{K}_{\parallel}\right)  \chi^{\mu}\left(  E\right)
\label{22}%
\end{equation}
with
\begin{equation}
W_{l}^{\mu}\left(  \mathbf{K}_{\parallel}\right)  =\frac{1}{2\pi}%
\int\mathrm{d}^{2}\mathbf{r}_{\parallel}\,\exp\left(  -i\mathbf{K}_{\parallel
}\mathbf{r}_{\parallel}\right)  \left[  \Omega_{l}^{\mu}\left(  \mathbf{r}%
_{\parallel}\right)  -\eta_{l}^{\mu}\right]  \label{23}%
\end{equation}
and, finally, the energy-dependent off-specular field in the Fraunhofer
approximation is
\begin{equation}
\Psi_{\mathrm{off}}\left(  \mathbf{r,}E\right)  =\sqrt{\frac{\pi}{2}}%
\frac{k^{2}}{r}\exp\left(  ikr\right)  \sum_{l,\mu}W_{l}^{\mu}\left(
\mathbf{K}_{\parallel}\right)  \chi^{\mu}\left(  E\right)  T_{l}\left(
E\right)  \Psi^{\mathrm{in}} \label{24}%
\end{equation}
while the off-specular intensity is
\begin{equation}
I_{\mathrm{off}}\left(  E\right)  =\frac{k^{4}}{4r^{2}}\sum_{ll^{\prime}\mu
\mu^{\prime}}C_{ll^{\prime}}^{\mu\mu^{\prime}}\left(  \mathbf{K}_{\parallel
}\right)  \operatorname{Tr}\left[  \Gamma_{l}^{\mu}\left(  E\right)
^{\dagger}\Gamma_{l^{\prime}}^{\mu^{\prime}}\left(  E\right)  \rho\right]
\text{.} \label{25}%
\end{equation}
The matrices for the homogeneous region $\mu$, $l$%
\begin{equation}
\Gamma_{l}^{\mu}\left(  E\right)  =\chi^{\mu}\left(  E\right)  T_{l}\left(
E\right)  \label{26}%
\end{equation}
are the products of the homogeneous solution $T_{l}\left(  E\right)  $ and the
susceptibility $\chi^{\mu}\left(  E\right)  $ of that region. For the sake of
brevity, henceforth the dependence of $T_{l}$ and $\Gamma_{l}^{\mu}$ on both
$k_{\perp}$ and $k_{\perp}^{\prime}$will not be explicitely written.

As we already indicated in the previous section, Eq. (\ref{18}) and,
consequently, Eq. (\ref{25}) differ from the usual expression of diffuse
intensity of DWBA as published earlier in the
literature\cite{Sinha88,Toperverg01a,Toperverg01b,Lauter94,Ruhm99} in such a
way, that the "distortions" are only considered on the incident path before
scattering, the exit path is left undistorted. The DWBA expression is given
and discussed in Appendix \ref{Appendix3}. Physically speaking, the difference
between the two approximations is that DIWA only takes distortions into
account on the incident wave, while DWBA on both incident and emerging waves.
Therefore a DIWA calculation is considerable faster but it is not expected to
be invariant with respect to exchanging the source and detector (a condition
widely called 'reciprocity'\cite{BornWolf,SCHIFF,Potton2004}). Should,
nevertheless, reciprocity be physically justified under some conditions, we
may take advantage of calculating only half of the '$\theta_{\mathrm{in}}-$
$\theta_{\mathrm{out}}$' and the '$\omega-2\theta$' maps along with mirroring
one side of the $\theta_{\mathrm{in}}=$ $\theta_{\mathrm{out}}$ line and the
$\omega=\theta$ lines onto the other, respectively. We shall see that the DIWA
calculation is quite accurate at one side of these lines, a fact finally
resulting both in saving another $50~\%$ of the computation time and in a high
accuracy of the calculation of the full diffuse scatter.

The geometrical (or 'domain') correlation function $C_{ll^{\prime}}^{\mu
\mu^{\prime}}~$between layers $l$\ and $l^{\prime}$and homogeneous parts $\mu
$\ and $\mu^{\prime}$of the layers with definition%
\begin{equation}
C_{ll^{\prime}}^{\mu\mu^{\prime}}\left(  \mathbf{K}_{\parallel}\right)
=\left(  2\pi\right)  W_{l}^{\mu}\left(  \mathbf{K}_{\parallel}\right)
^{\ast}\ W_{l^{\prime}}^{\mu^{\prime}}\left(  \mathbf{K}_{\parallel}\right)
\label{27}%
\end{equation}
may describe quite general structural and magnetic intra- and interlayer
correlations like correlated layer growth, or magnetic-magnetic correlation
(e.g. antiferromagnetic multilayer domains) and even magnetic-structural
correlations. Similarly to Eq. (\ref{17}), the direct space correlation
function can be written as
\begin{equation}
{\mathfrak{C}}_{ll^{\prime}}^{\mu\mu^{\prime}}\left(  \mathbf{R}_{\parallel
}\right)  =\frac{1}{A}\int\mathrm{d}^{2}\mathbf{r}_{\parallel}\,\left[
\Omega_{l}^{\mu}\left(  \mathbf{R}_{\parallel}+\mathbf{r}_{\parallel}\right)
-\eta_{l}^{\mu}\right]  \left[  \Omega_{l^{\prime}}^{\mu^{\prime}}\left(
\mathbf{r}_{\parallel}\right)  -\eta_{l^{\prime}}^{\mu^{\prime}}\right]  .
\label{27B}%
\end{equation}
Notice that both the correlation function and its Fourier transform are
symmetric with respect to the simultaneous exchange of the domain and layer
indices:%
\begin{equation}
{\mathfrak{C}}_{ll^{\prime}}^{\mu\mu^{\prime}}\left(  \mathbf{R}_{\parallel
}\right)  ={\mathfrak{C}}_{l^{\prime}l}^{\mu^{\prime}\mu}\left(
\mathbf{R}_{\parallel}\right)  . \label{27C}%
\end{equation}
An important consequence of Eqs. (\ref{19p}) and (\ref{21a}) is
\begin{equation}%
{\displaystyle\sum\limits_{\mu^{\prime}}}
{\mathfrak{C}}_{ll^{\prime}}^{\mu\mu^{\prime}}\left(  \mathbf{R}_{\parallel
}\right)  =%
{\displaystyle\sum\limits_{\mu}}
{\mathfrak{C}}_{ll^{\prime}}^{\mu\mu^{\prime}}\left(  \mathbf{R}_{\parallel
}\right)  =0. \label{27D}%
\end{equation}
\ Using Eq. (\ref{27B}) the correlation function at $\mathbf{R}_{\parallel}=0$
reads
\begin{equation}
{\mathfrak{C}}_{ll^{\prime}}^{\mu\mu^{\prime}}\left(  0\right)  =\eta
_{ll^{\prime}}^{\mu\mu^{\prime}}-\eta_{l}^{\mu}\eta_{l^{\prime}}^{\mu^{\prime
}} \label{27E}%
\end{equation}
where
\begin{equation}
\eta_{ll^{\prime}}^{\mu\mu^{\prime}}=\frac{1}{A}\int\mathrm{d}^{2}%
\mathbf{r}_{\parallel}\,\Omega_{l}^{\mu}\left(  \mathbf{r}_{\parallel}\right)
\Omega_{l^{\prime}}^{\mu^{\prime}}\left(  \mathbf{r}_{\parallel}\right)
\label{27F}%
\end{equation}
is the fractional overlap of homogeneous parts of types $\mu$ and $\mu
^{\prime}$ in layers $l$ and $l^{\prime}$,\ respectively.

In Eq. (\ref{25}) the geometrical correlation is separated from the
energy-dependence and, therefore, it can be applied for time and energy domain
experiments alike. The Fourier transformation can be performed so
that,\emph{\ }using Eqs. (\ref{19}) and (\ref{25}), the time-dependent
intensity becomes
\begin{equation}
I_{\mathrm{off}}\left(  t\right)  =\frac{k^{4}}{4r^{2}}\sum_{ll^{\prime}\mu
\mu^{\prime}}C_{ll^{\prime}}^{\mu\mu^{\prime}}\left(  \mathbf{K}_{\parallel
}\right)  \operatorname{Tr}\left[  G_{l}^{\mu}\left(  t\right)  ^{\dagger
}\ G_{l^{\prime}}^{\mu^{\prime}}\left(  t\right)  \rho\right]  \label{28}%
\end{equation}
where%
\begin{equation}
G_{l}^{\mu}\left(  t\right)  =\frac{1}{\hbar\sqrt{2\pi}}\int\limits_{-\infty
}^{\infty}\mathrm{d}E\,\Gamma_{l}^{\mu}\left(  E\right)  \exp\left(
-iEt/\hbar\right)  . \label{29}%
\end{equation}

\subsection{Time-integral scattered intensity}

The time-integrated intensity is $I_{\mathrm{off}}^{\mathrm{int}}%
=\int\limits_{t_{1}}^{t_{2}}\mathrm{d}t\,I_{\mathrm{off}}\left(  t\right)  $
where $t_{1}$ and $t_{2}$ define the starting and finishing time of the time
window of the counting after the synchrotron pulse. Applying Eqs. (\ref{28})
and (\ref{29})
\begin{equation}
I_{\mathrm{off}}^{\mathrm{int}}=\frac{k^{4}}{4\hbar r^{2}}\sum_{m=-\infty
}^{\infty}s_{m}\int\limits_{-\infty}^{\infty}\mathrm{d}E\sum_{ll^{\prime}%
\mu\mu^{\prime}}C_{ll^{\prime}}^{\mu\mu^{\prime}}\left(  \mathbf{K}%
_{\parallel}\right)  \operatorname{Tr}\left[  \Gamma_{l}^{\mu}\left(
E+m\varepsilon\right)  ^{\dagger}\ \Gamma_{l^{\prime}}^{\mu^{\prime}}\left(
E\right)  \rho\right]  \label{30}%
\end{equation}
where $\varepsilon=\frac{h}{t_{\mathrm{bunch}}}$ with $t_{\mathrm{bunch}}$
being the time interval between the synchrotron bunches, $h$ is the Planck
constant, $s_{m}$ is the $m^{\mathrm{th}}$ discrete Fourier component of the
periodical time window function $S\left(  t\right)  =\sum\limits_{m=-\infty
}^{\infty}s_{m}\exp\left(  i\frac{2m\pi}{t_{\mathrm{bunch}}}t\right)  $ of the
experiment defined by $S\left(  t\right)  =1$ for $t_{1}<t<t_{2}$, otherwise
$S\left(  t\right)  =0$ after each synchrotron bunch.

\section{Model calculations and comparison with experiment\label{Applications}%
}

Applying the above theory, off-specular time-integrated SMR curves and maps
were calculated and, on one hand, compared with PNR curves and maps simulated
with the same DIWA theory and, on the other hand, compared with experimental
off-specular SMR data. In order to treat realistic problems, at this point
further specification of the studied system\emph{\ }and the\emph{\ }%
experimental\emph{\ }conditions\emph{\ }will be undertaken\emph{\ }--- without
restricting$~$the$~$generality~of\emph{~}the discussion.\emph{\ }%
The\emph{\ }actual\emph{\ }calculations\emph{\ }were\emph{\ }%
performed\emph{\ }for\emph{\ }the\emph{\ }$\mathrm{MgO}/\left[  ^{57}%
\mathrm{Fe}\left(  2.62\,\mathrm{nm}\right)  /\mathrm{Cr}\left(
1.28\,\mathrm{nm}\right)  \right]  _{20}$\emph{\ }multilayer\emph{\ }%
structure.\emph{\ }No interface layers and no roughness contribution were
considered. Although, as mentioned,\emph{\ }the\emph{\ }$C_{l^{\prime}l}%
^{\mu^{\prime}\mu}$ functions may describe a variety of domain correlations,
simulations\emph{\ }were\emph{\ }performed\emph{\ }for\ the case
of\emph{\ }strongly-coupled layer antiferromagnet in remanence.\emph{\ }%
By\emph{\ }layer antiferromagnet we mean even number of magnetic layers of
identical thickness resulting in zero net magnetization of the ML stack in
remanence.\emph{\ }Except for the trivial case\emph{\ }of full in-plane
saturation, the magnetic layers in an AF multilayer stack are broken into
domains of different orientations. In remanence, we assume $180^{\circ}%
$\ domain walls of negligible thickness, i.e. the sublayer magnetizations also
vanish (for further details see below). Strong coupling implies a strict
plane-perpendicular domain correlation throughout the ML stack, a structure,
in which the top layer unequivocally identifies the domain structure in the
lower ones (say\ '$+$' and\ '$-$' type domains). The actual functional form of
such AF domain correlation is given in the next section.

For the SMR simulations, the scattering geometry was selected so that the
layer magnetizations lay both in the plane of the film and of scattering, i.e.
parallel/antiparallel to the in-plane component of the wave vector,
$\mathbf{k}_{\parallel}$, a condition for the appearance of the SMR specular
AF reflection.\cite{Chumakov99} All SMR curves and maps were calculated for
the 14.4-keV M\"{o}ssbauer resonance of $^{57}\mathrm{Fe}$ ($\lambda
=0.086~\mathrm{nm}$) for hyperfine magnetic fields of $\pm33.08~\mathrm{T}$
for the'$+$' and\ '$-$' type domains, respectively. The isomer shift and
quadrupole splitting were set to zero (parameters for \textit{bcc} iron at
room temperature). The electronic susceptibilities for the $\lambda
=0.086~\mathrm{nm}$ x-rays were taken from the Berkeley web site\cite{Henke}
for the various elements. The synchrotron bunch time $t_{\mathrm{bunch}}$, and
integration boundaries $t_{1}$ and $t_{2}$ were chosen according to the actual
experimental values (see below).

For PNR simulations, monochromatic neutrons of wavelength $\lambda
=0.4~\mathrm{nm}$ were assumed and the layer magnetizations were set
parallel/antiparallel to the neutron spin and perpendicular to $\mathbf{k}%
_{\parallel}$. The nuclear scattering lengths used for Cr, $^{57}$Fe and MgO
are 3.6, 2.3, and $11.2~\mathrm{fm}$, respectively.\cite{Neutron News} The
magnetic scattering length $b_{\mathrm{m}}(z)=$ $C\mu(z)$, where
$C=r_{0}\gamma/2=2.69542~\mathrm{fm}/\mu_{\mathrm{B}}$, where $r_{0}$ is the
classical electron radius, $\gamma=1.91304$ is the magnetic moment of neutron
in nuclear magnetons, $\mu(z)$ is the average magnetic moment per \ atom per
unit volume at depth $z$,$~\mu_{\mathrm{B}}$ is the Bohr magneton.
Numerically, $b_{\mathrm{m}}(z)$ was set to $\pm5.93~\mathrm{fm}$ for the
'$+$' and\ '$-$' type domains within the $^{57}$Fe layers and zero otherwise.

The theory presented above was implemented in, and specular and off-specular
intensity curves and maps (both SMR and PNR)\emph{\ }were simulated by the
data evaluation computer program EFFI (\textbf{E}nvironment \textbf{F}or
\textbf{FI}tting), which is freely downloadable.\cite{Spiering00,EFFI}

\subsection{Consequences of the finite coherence length}

In the case of a 1:1 surface coverage of the '$+$'\ and\ '$-$'\ type domains,
laterally averaging the magnetizations within a layer --- as mentioned in the
introduction --- leads to a loss of the AF contrast and no AF Bragg peak in
the specular reflectivity appears. However, when the lateral size of the
domains is bigger than or comparable to the lateral coherence
length\cite{Toperverg01a}\ of the applied radiation, a net layer magnetization
is sampled within the coherence area and such intensity contributions are to
be added incoherently.\cite{Baron96,Sinha1998,Toperverg01a} Consequently, a
magnetic contrast appears in the specular reflectivity. In order to account
for the effects of the finite coherence volume,\cite{Mandel-Wolf} the domain
size is to be related to the lateral coherence area, the projection of the
coherence volume\cite{Mandel-Wolf} to the top magnetic layer. (The lateral
coherence length was reported to be 0.1 to $30\ \mathrm{\mu m~}$for
neutrons\cite{Paul} and similar values can be derived for nuclear resonant
photons.\cite{Baron96})

In order to account for the effects of the finite coherence
volume,\cite{Mandel-Wolf} the domain size is to be related to the coherence
area. In order to account for the effect of the finite coherence length we
redefine the fractional area of a domain type, $\eta_{l}^{\mu}$ that was
defined above for incident radiation with infinite coherence lengths. Indeed,
as a first approximation, one can use Eq. (\ref{defeta}) inside the coherence
area within the \ top layer by exchanging $A$, the total area of the film, for
the coherence area and relating the surface of type $\mu$ domains to the
coherence area. In case of an antiferromagnet the domain type index $\mu$
identifies the '$+$'\ and\ '$-$'\ type domains ($\mu=+,-$). Using Eq.
(\ref{21a}) one can introduce a specific magnetic bias parameter\ $\eta$\ for
the layers of even and odd index by the definition%
\begin{equation}
\eta=\eta_{2i+1}^{+}=1-\eta_{2i+1}^{-}=\eta_{2i}^{-}=1-\eta_{2i}^{+}
\label{af_eta}%
\end{equation}
with $i$ being an integer.\ 

\begin{figure}[p]
\includegraphics{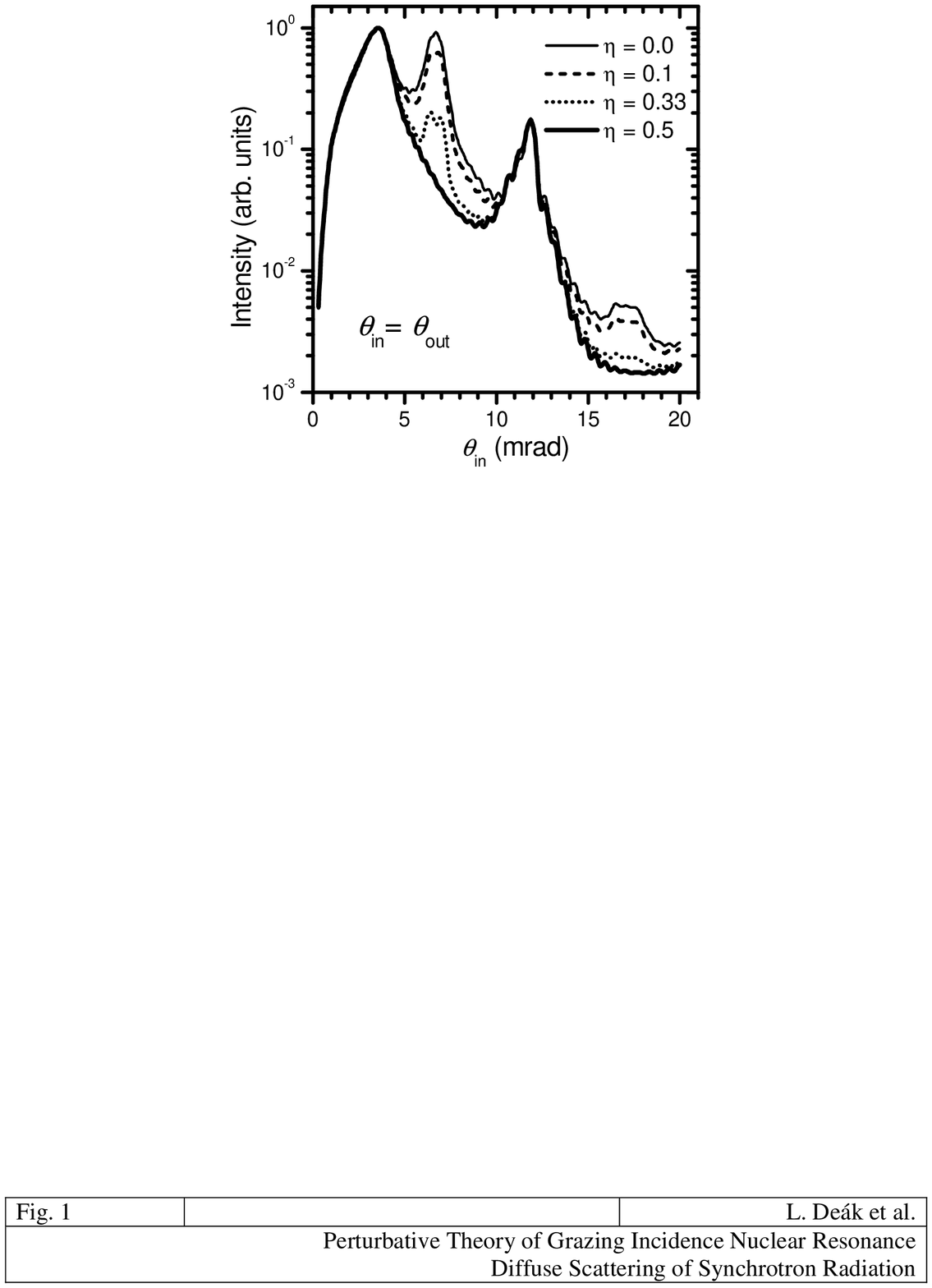}
\caption{Simulated $^{57}\mathrm{Fe}$ specular synchrotron M\"{o}ssbauer
reflectograms ($\theta-2\theta$ scans) of the $\mathrm{MgO}/\left[
^{57}\mathrm{Fe}\left(  2.62\,\mathrm{nm}\right)  /\mathrm{Cr}\left(
1.28\,\mathrm{nm}\right)  \right]  _{20}$ antiferromagnetic multilayer with
sublayer magnetizations parallel and antiparallel to the wave vector for
different domain bias parameters, $\eta$ indicated in the figure.}%
\label{spec-smr}%
\end{figure}

In Fig.\ \ref{spec-smr}\ simulated specular SMR intensity curves are displayed
for different values of\ $\eta$. As expected, the $\eta$-dependence is
restricted to the $\theta_{\text{\textrm{in}}}$ $\left(  =\theta
_{\text{\textrm{out}}}\right)  $ regions of $1/2$- and $3/2$-order AF Bragg
reflections. As explained, there are no AF peaks for\ $\eta=0.5$ and the
magnetic contrast increases with increasing dominance of either '$+$%
'\ or\ '$-$'\ type domains.\ Since the coherence area is smaller then the
illuminated area and it may have an arbitrary position, which can be
identified by $\eta$, the integration in Eq. (\ref{17}) is performed as a
function of $\eta$.\ The off-specular intensity in (\ref{18}) is calculated as%
\begin{equation}
I_{\mathrm{off}}=%
{\displaystyle\int\limits_{0}^{1}}
p\left(  \eta\right)  I_{\mathrm{off}}\left(  \eta\right)  \mathrm{d}\eta,
\label{coh_integral}%
\end{equation}
where $p\left(  \eta\right)  $\ is the normalized probability density of
having a (magnetic) domain bias of $\eta$. We note that the probability
function $p\left(  \eta\right)  $ depends both on the lateral domain structure
and on the measurement geometry (source and slit dimensions, etc.). For the
sake of simplicity, we consider here the case when the probability density
$p\left(  \eta\right)  $\ sharply peaks near $0$\ and $1$ since strong
specular SMR\ AF Bragg peaks can indeed be observed experimentally. Since the
NRS spectra of a magnetic layer with $k$-parallel or a $k$-antiparallel
directions of the hyperfine field can not be distinguished, the SMR AF\ Bragg
peak is of the same shape and intensity irrespective of whether the AF
structure starts with a $k$-parallel or a $k$-antiparallel layer on the top of
the stack.\cite{Nagy99} Therefore the curves in Fig. \ref{spec-smr} are
identical for domain bias\ $\eta$ and $1-\eta$.\ 

The specular curves in Fig. \ref{spec-smr} are typical for an AF multilayer.
For the above model structure and wavelength corresponding to the $^{57}$Fe
M\"{o}ssbauer resonance energy the structural and AF Bragg peaks show up at
$\theta_{\mathrm{in}}=11.3\,\ $and $\theta_{\mathrm{in}}=6.7\,$mrad,
respectively. The Kiessig fringes,\cite{Kiessig31} characteristic of the total
thickness of the ML, appear in all specular curves, however, due to the strong
nuclear resonant absorption in the $^{57}$Fe layers, their amplitude is
strongly attenuated. Below the critical angle, the delayed reflected SMR
intensity tends to zero at zero incident angle,\cite{Deak94,Baron94a} whereas
a non-resonant x-ray or PNR curve behaves 'normally' (i.e. tends to unity,
corresponding to total reflection). This difference, as we shall see, leads to
an augmented effect on the diffuse intensities.

\subsection{Lateral correlation: domains}

The domain structure within the top layer of the AF\ multilayer stack will be
assumed laterally isotropic and statistically characterized by the correlation
function of the top layer magnetization. For simplicity, we assume an
exponential function and interpret the average domain size as the correlation
length $\xi$ of the exponential correlation function\cite{Sinha88}%
\begin{equation}
{\mathfrak{C}}_{ll^{\prime}}^{\mu\mu^{\prime}}\left(  \mathbf{r}_{\parallel
}\right)  =\left(  -1\right)  ^{l+l^{\prime}+1-\delta_{\mu\mu^{\prime}}}%
\ \eta\left(  1-\eta\right)  \exp\left(  -\frac{r_{\parallel}}{\xi}\right)
\label{35}%
\end{equation}
where $\delta$ is the Kronecker delta symbol. We assume the same correlation
function for each values of indices observing an alternating sign
corresponding to the strict plane-perpendicular AF correlation and an in-plane
'$+$'/'$-$' domain model. The Fourier transform of Eq. (\ref{35}) reads
\begin{equation}
C_{ll^{\prime}}^{\mu\mu^{\prime}}\left(  \mathbf{K}_{\parallel}\right)
=\left(  -1\right)  ^{l+l^{\prime}+1-\delta_{\mu\mu^{\prime}}}\ \eta\left(
1-\eta\right)  \frac{2\pi\xi^{2}}{\left[  1+\left(  K_{\parallel}\,\xi\right)
^{2}\right]  ^{3/2}}, \label{36}%
\end{equation}
which is then substituted into Eqs. (\ref{25}) and (\ref{30}) for calculating
the off-specular PNR and SMR intensities, respectively.

\subsection{\bigskip Off-specular scattering: PNR\ and SMR}

\begin{figure}[p]
\includegraphics{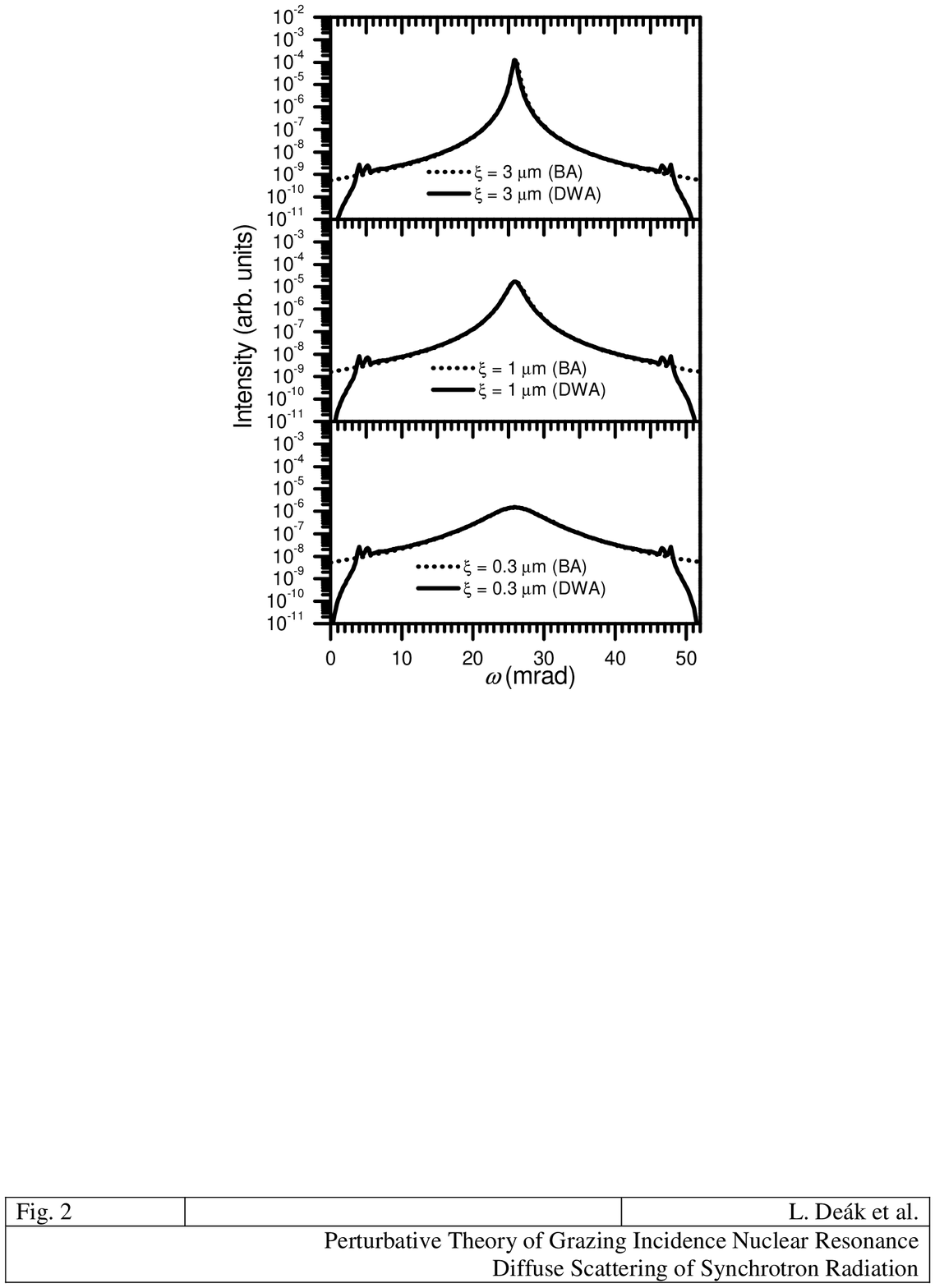}
\caption{Simulated off-specular PNR $\omega-$scans ($I^{++}$) calculated for
the $\mathrm{MgO}/\left[  ^{57}\mathrm{Fe}\left(  2.62\,\mathrm{nm}\right)
/\mathrm{Cr}\left(  1.28\,\mathrm{nm}\right)  \right]  _{20}$
antiferromagnetic multilayer with $\lambda=0.4$ nm\ and detector position
$2\theta$ fixed at the $1/2$-order antiferromagnetic Bragg peak position for
various correlation lengths, $\xi$ indicated in the figure. A single domain
bias parameter of $\eta=0.1$ was used. Dotted and solid lines show the BA and
SDIWA simulations, respectively.}%
\label{omega_pnr}%
\end{figure}

\begin{figure}[p]
\includegraphics{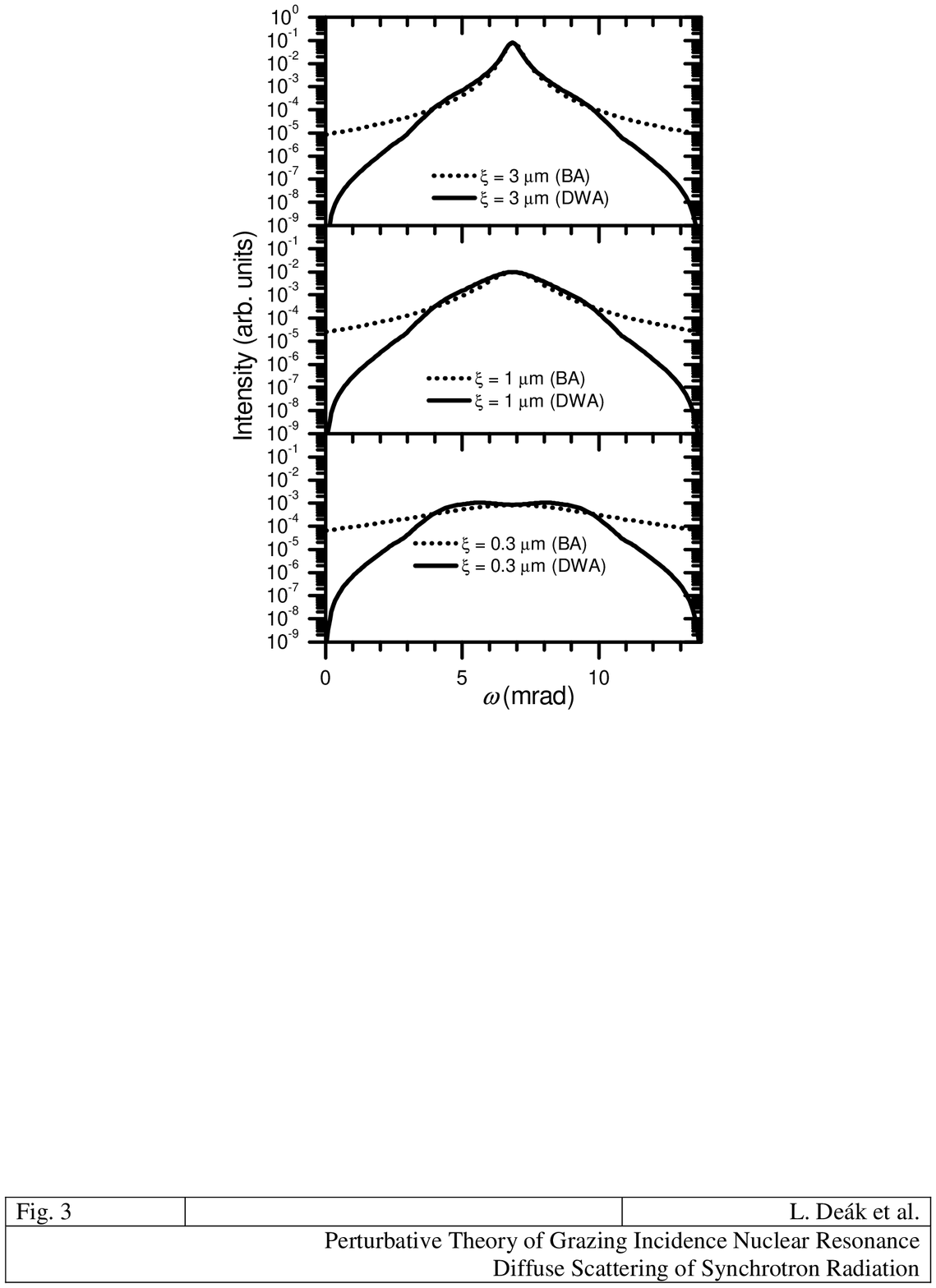}
\caption{Simulated off-specular SMR $\omega-$scans calculated for the
$\mathrm{MgO}/\left[  ^{57}\mathrm{Fe}\left(  2.62\,\mathrm{nm}\right)
/\mathrm{Cr}\left(  1.28\,\mathrm{nm}\right)  \right]  _{20}$
antiferromagnetic multilayer with $\lambda=0.086$ nm\ of the $^{57}$Fe
M\"{o}ssbauer radiation and the detector position $2\theta$ fixed at the
$1/2$-order antiferromagnetic Bragg peak position for various correlation
lengths, $\xi$ indicated in the figure. A single domain bias parameter of
$\eta=0.1$ was used. Dotted and solid lines show the BA and SDIWA simulations,
respectively.}%
\label{omega_smr}%
\end{figure}

Figs. \ref{omega_pnr} and \ref{omega_smr} show simulated $\omega$-scans for
the above model multilayer at $2\theta=52~$mrad and $2\theta=13.4~$mrad
corresponding to the $1/2$-order AF\ Bragg angle for different domain
correlation lengths $\xi$ using a single domain bias parameter $\eta=0.1$, for
PNR and SMR, respectively. Applying Eqs. (\ref{C6})-(\ref{C9}) the kinematical
approximation (first Born approximation, BA) is also shown. The curves were
symmetrized by substituting the mirror image of $\omega<\theta~$in place of
the $\omega>\theta$\ region. The grounds of such substitution will be explaned
below. The BA curves display the same shape in the entire $\xi$-range and the
off-specular scatter width is inversely proportional to $\xi$. The BA and
symmetrized DIWA (called, henceforth, 'Symmetrized Distorted Incident-Wave
Approximation', SDIWA) curves for PNR overlap almost in the entire angular
range, the differences only develop near the critical incident and exit
angles, i.e. at the Yoneda wings. For SMR, the SDIWA curves change their shape
for different correlation lengths, since, as expected, the plane-perpendicular
electronic and hyperfine depth profile, the energy- and polarization-dependent
absorption embedded in the $\Gamma_{l}^{\mu}$\ functions have a strong
influence on the shape of the off-specular scatter. Unlike in case of PNR,
therefore,\emph{\ }identifying the correlation length with the inverse width
of the diffuse SMR scatter -- without a proper evaluation of the entire
reflectivity curve -- can not be justified. Note, that the simulations were
performed for the experimentally feasible neutron and nuclear photon
wavelengths of $\lambda_{n}=0.4~\mathrm{nm}~$and $\lambda_{^{57}\mathrm{Fe}%
}=0.086~\mathrm{nm}$,~respectively. Although the scattering amplitudes are
comparable, the angular range between the Yoneda wings, consequently the
angular width of validity of the BA approximation is considerably wider for
the longer wavelength neutrons. This, however, does not effect the above
statements on the qualitative differences between the PNR\ and SMR maps.

\begin{figure}[p]
\includegraphics{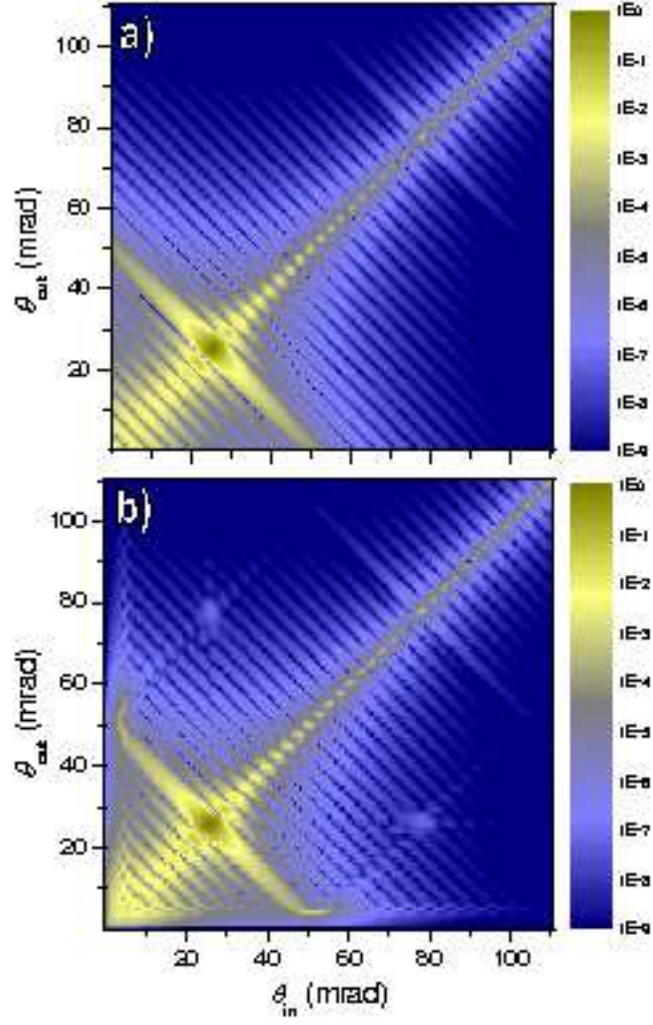}
\caption{Simulated '$\theta_{\mathrm{in}}-\theta_{\mathrm{out}}$' PNR\ diffuse
intensity maps for the $\mathrm{MgO}/\left[  ^{57}\mathrm{Fe}\left(
2.62\,\mathrm{nm}\right)  /\mathrm{Cr}\left(  1.28\,\mathrm{nm}\right)
\right]  _{20}$ antiferromagnetic multilayer structure. The intensities are
normalized and shown on a logarithmic color scale. BA (a) and SDIWA (b)
$I^{++}$ intensities are shown. A single domain bias parameter of $\eta=0.1$
was used.}%
\label{twodfull-pnr}%
\end{figure}

\begin{figure}[p]
\includegraphics{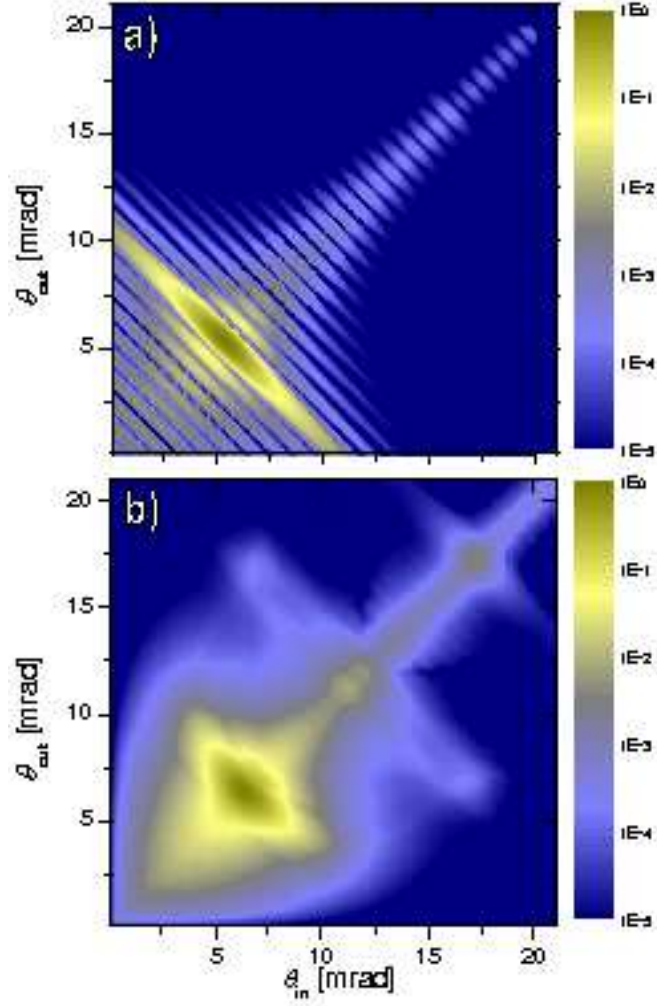}
\caption{Simulated '$\theta_{\mathrm{in}}-\theta_{\mathrm{out}}$' SMR diffuse
intensity maps of the $\mathrm{MgO}/\left[  ^{57}\mathrm{Fe}\left(
2.62\,\mathrm{nm}\right)  /\mathrm{Cr}\left(  1.28\,\mathrm{nm}\right)
\right]  _{20}$ antiferromagnetic multilayer structure around the $1/2$-order
antiferromagnetic Bragg peak using the BA (a) and the SDIWA (b). The
intensities are shown on a logarithmic color scale and are normalized. A
single domain bias parameter of $\eta=0.1$ was used.}%
\label{twod-smr}%
\end{figure}

Figs. \ref{twodfull-pnr}b\ and \ref{twod-smr}b display simulated
two-dimensional '$\theta_{\mathrm{in}}-\theta_{\mathrm{out}}$' maps for
$\xi=1~\mathrm{\mu m}$ correlation length with a single domain bias parameter
$\eta=0.1$ in the same DW approximation, corresponding to Eqs. \ref{25} and
\ref{30} for PNR\ and SMR, respectively. The intensity is maximum along the
diagonal specular line, and a broad diffuse intensity is observed around the
half-order Bragg peaks. Similarly to the $\omega$-scans, the\emph{\ }%
$\theta_{\mathrm{in}}<\theta_{\mathrm{out}}$\emph{\ }semi-plane was mirrored
onto the\emph{\ }$\theta_{\mathrm{in}}>\theta_{\mathrm{out}}$\emph{\ }%
semi-plane. Only the diffuse $I^{++}$ SDIWA and BA maps are displayed for PNR
in Fig. \ref{twodfull-pnr}. $I^{+-}$and $I^{-+}$ maps are not shown since the
spin-flip scattering vanishes due to the '$+$'/'$-$' domains being assumed
parallel/antiparallel with the incident neutron spin. Moreover, due to the
small absorption of the neutrons, the $++$ and $--$ maps (corresponding to the
same layer structure with reversed layer sequence), are practically identical
(and therefore not shown). It is not surprising that the Kiessig fringes are
observed in both PNR\ and SMR diffuse scatter since the source for the diffuse
intensity is the specular field. Due to the negligible absorption of the
neutrons, the Kiessig contrast is stronger in PNR than in SMR. We may observe
further symmetries in the PNR\ curve. As expected, the diffuse intensity
around the structural Bragg node at $\theta_{\mathrm{in}}=\theta
_{\mathrm{out}}=52~\mathrm{mrad}$ is missing, since the diffuse scattering
here is purely magnetic origin and the magnetic contributions cancel each
other at the momentum transfer corresponding to the first order Bragg peak.
Moreover, for $^{57}$Fe/Cr for a 2:1 layer thickness ratio the 3/2-order Bragg
reflection is forbidden, consequently, the node for the 2.62~nm($^{57}%
$Fe)/1.28~nm(Cr) layer thickness ratio in Figs. \ref{twodfull-pnr},\ and
\ref{twod-smr} are very weak.

The $^{57}$Fe $\theta_{\mathrm{in}}-\theta_{\mathrm{out}}$ SMR map in Fig.
\ref{twod-smr}b shows a number of features different from that of PNR. Due to
the dominance of energy-dependent absorption for resonant x-rays the SMR BA
and SDIWA maps drastically differ. E.g., weak structural Bragg wings
around\emph{\ }$\theta_{\mathrm{in}}=\theta_{\mathrm{out}}%
=11.25\ \mathrm{mrad}$\emph{\ }appears, but those fall below the typical
experimental background level. Below the critical angle (unlike in cases of
PNR or non-resonant x-rays), the \textit{specular} SMR intensity, the source
of the diffuse scatter, rapidly decreases to zero (\textit{cf}. Fig.
\ref{spec-smr}). Since, independently, due to the stronger absorption, the
Kiessig fringes are suppressed in the SMR map in Fig. (\ref{twod-smr}b), the
intensity does not oscillate near Yoneda wings, which, as a result, appear
broadened as compared to the PNR pattern in Fig. (\ref{twodfull-pnr}b). As a
consequence, the total-reflection peak, being in the critical region, is
somewhat difficult to distinguish from the $1/2-$order AF Bragg node in Fig.
(\ref{twod-smr}b).

So far, by discussion of simulated SMR and PNR-scans and maps, we have shown,
that the presented DIWA\ approach satisfactorily describes both the PNR\ and
SMR\ off-specular intensity --- except for \textit{exit }angles below and
around the critical angle of total reflection. In Appendix \ref{Appendix3} the
more exact DWBA SMR intensity formula is also given, which takes into account
the distortions on both incident and emerging waves. However, the typical
computation time needed for the DWBA calculation renders its usage completely
impractical for fitting of diffuse synchrotron M\"{o}ssbauer reflectograms
even on advanced present-day computer architectures. Indeed, Eqs. (\ref{25})
and (\ref{C6}), differ only in the definition of $\Gamma_{l}^{\mu}$ according
to Eqs. (\ref{26}) and (\ref{C7}), for DIWA and for DWBA, respectively.
Counting the number of the $2\times2$ complex matrix multiplications one can
compare the speed of the two algorithms. On the one hand, in the DIWA
equations (\ref{26}) and (\ref{B7}), one has to perform nine complex
$2\times2$ matrix multiplications. On the other hand, in the DWBA Eq.
(\ref{C7}) $N_{p}\times13$ complex $2\times2$ matrix multiplication have to be
performed. $N_{p}$ is typically ten. One may, therefore, conclude that DIWA,
as derived here, is, at least, $10\times13/9\approx14$ times faster, than
DWBA.\ Since, as previously explained, a single diffuse SMR\ map requires as
much more computing time as the number of energy channels in the M\"{o}ssbauer
spectrum (typically 1024), this further factor in computing time\ would make
such calculations extremely tedious.\cite{Machine-time} Therefore an
alternative approach was followed here.

In case of PNR, as a consequence of the negligible absorption of neutrons
(Hermitian scattering potential), reciprocity is fulfilled. Conversely, the
resonant M\"{o}ssbauer medium is absorptive and gyrotropic and reciprocity can
not be proved to hold in general.\cite{Potton2004,Huffman1970} In fact, the
violation of reciprocity in scattering on absorptive and grotropic media is
still the subject of both
experimental\cite{Chernov2003,Chernov2000,Chernov2005} and
theoretical\cite{Potton2004,Mytnichenko2005} works.\cite{reciprocity}

Nevertheless, for the special case under discussion, when all hyperfine
magnetic fields are aligned parallel/antiparallel to the wave vector, one may
verify that reciprocity exactly holds. Indeed, in this case, the scattering
amplitude is diagonal on the circular polarization basis throughout the whole
multilayer and, therefore, Eq. (\ref{1}) becomes uncoupled. Consequently the
conventional proof of reciprocity to both eigenmodes is
straightforward\cite{SCHIFF,BornWolf} and the '$\theta_{\mathrm{in}}-$
$\theta_{\mathrm{out}}$' SMR maps become symmetrical. In view of the above, we
presribe reciprocity for both PNR and SMR and mirror the $\theta_{\mathrm{in}%
}<\theta_{\mathrm{out}}$ semi-plane onto the $\theta_{\mathrm{in}}%
>\theta_{\mathrm{out}}$ semi-plane. By doing so, the DWBA accuracy of the
$\theta_{\mathrm{in}}<\theta_{\mathrm{out}}$ semi-plane is achieved by the
present (faster) DIWA algorithm on the entire '$\theta_{\mathrm{in}}-$
$\theta_{\mathrm{out}}$' plane (with a further decrease of computing time by a
factor of two. This symmetrizing procedure (SDIWA) was used to simulate the
DIWA curves and maps in this paper (cf. Figs. \ref{omega_pnr}, \ref{omega_smr}%
, \ref{twodfull-pnr}, \ref{twod-smr} and \ref{expsim2D}).

\subsection{Experimental results and
discussion\label{Experimental results and discussion}}

\begin{figure}[p]
\includegraphics{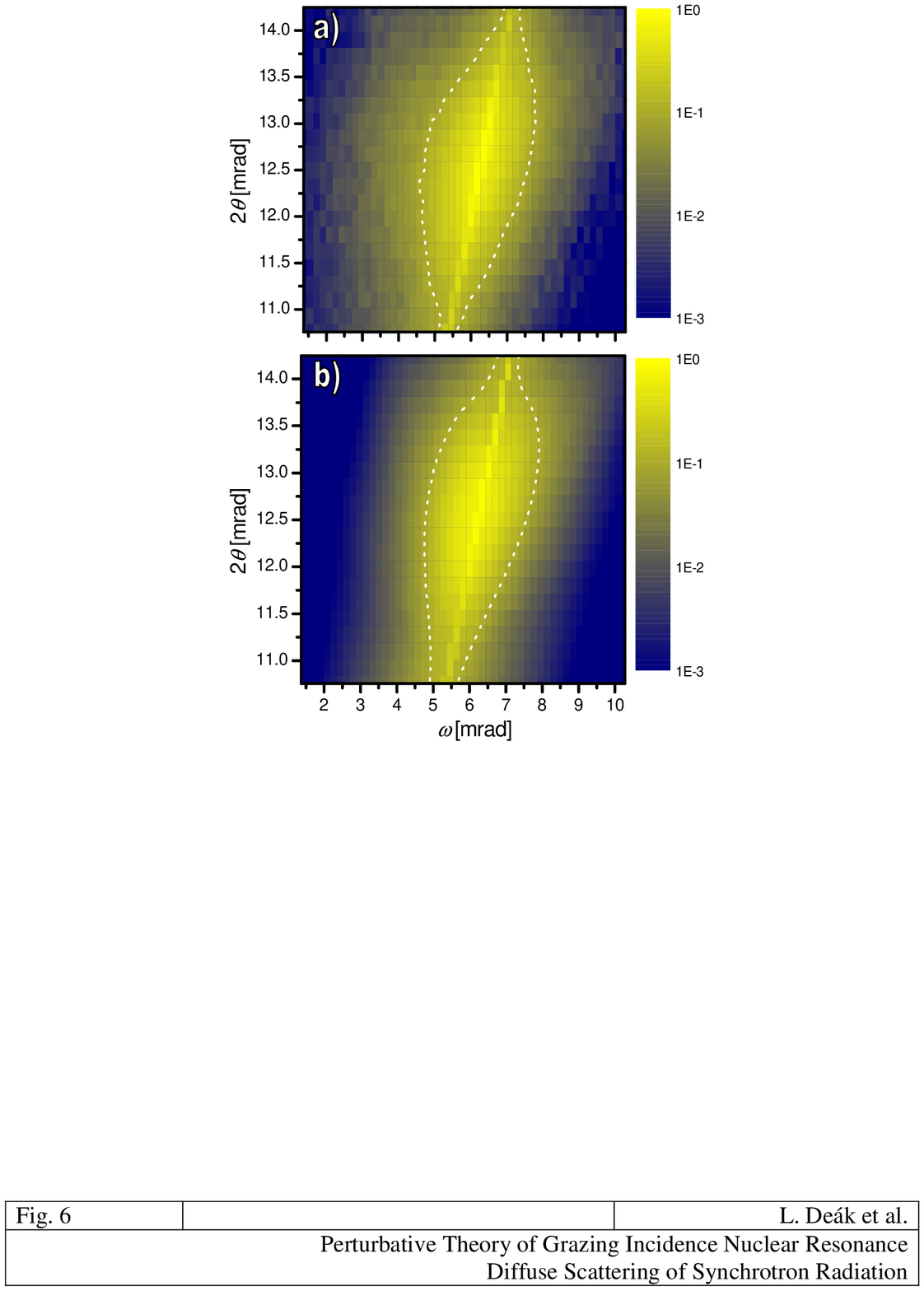}
\caption{Measured (a) and simulated (b) '$\omega-2\theta$' SMR diffuse
intensity maps of the $\mathrm{MgO}/\left[  ^{57}\mathrm{Fe}\left(
2.62\,\mathrm{nm}\right)  /\mathrm{Cr}\left(  1.28\,\mathrm{nm}\right)
\right]  _{20}$ antiferromagnetic multilayer in the vicinity of the
$1/2$-order antiferromagnetic Bragg peak. The intensities are normalized and
shown on a logarithmic color scale. Map (b) was simulated using Eqs. \ref{30},
\ref{35} and \ref{36} with $\xi=1.0~\mathrm{\mu m}$ domain correlation length.
The dotted white line shows the 10\% level of the maximum intensity.}%
\label{expsim2D}%
\end{figure}

Two-dimensional experimental and simulated '$\omega-2\theta$' SMR\ maps of a
[$^{57}$Fe/Cr] antiferromagnetic multilayer are presented in Fig.
\ref{expsim2D} in the vicinity of the antiferromagnetic ($1/2-$order) Bragg
reflection (region marked by dashed lines in Fig. (\ref{twod-smr})). The
MgO$(001)$/[$^{57}$Fe/Cr]$_{20}$ ML was prepared by molecular beam epitaxy at
the IMBL facility of IKS Leuven, Belgium. Preparation and characterization of
the sample has been described
elsewhere.\cite{BottyanBSF1,Nagy02a,Tancziko2004} The layering was verified
epitaxial and periodic, with thicknesses of $2.6~\mathrm{nm}$ and
$1.3~\mathrm{nm}$ for the $^{57}$Fe and Cr layers, respectively. SQUID
magnetometry showed a saturation field of $0.9~\mathrm{T}$ and AF coupling
between neighboring Fe layers. According to previous studies on this
multilayer,\cite{Nagy02a,BottyanBSF1,Tancziko2004} the Fe magnetizations at
remanence align along the $(100)$ and $(010)$ perpendicular easy directions
corresponding to the respective $(110)$ and $(\overline{1}10)$ directions of
the MgO substrate. Experimental realization of alignment of domains along the
$k$-vector was achieved in the following way. First a magnetic field
of\emph{\ }$1.6$ $\mathrm{T}$\emph{\ }was applied perpendicular to $k$\ in one
of the two equivalent in-plane directions of easy magnetization of the
MgO$(001)$/[$^{57}$Fe/Cr]$_{20}$ ML, then the field was relaxed to remanence.
By this\emph{\ }procedure, due to the antiferromagnetic coupling between the
layers, the sublayer magnetizations became aligned in the perpendicular easy
direction, the two types of AF domains being only different in the top-layer
magnetization direction.\cite{BottyanBSF1,Nagy02a,Nagy02b}

Experimental $\omega-2\theta$ SMR maps were recorded at the BL09XU nuclear
resonance beam line\cite{Yoda2001} of SPring-8, Japan, using the $14.4$
$\mathrm{keV}$ M\"{o}ssbauer transition of $^{57}$Fe by performing sequential
$\omega$-scans in a $2\theta$-range. The synchrotron was operated in the
203-bunch mode, corresponding to a bunch separation time of $t_{\mathrm{bunch}%
}=23.6\ \mathrm{ns}$. The SR was monochromatized by a Si$(4\,2\,2)$%
/Si$(12\,2\,2)$ double channel-cut high-resolution monochromator with a
resolution of $6~\mathrm{meV}$. The specimen was\emph{\ }mounted in
grazing-incidence geometry with a sample-to-detector distance of 46 cm and
receiving slit width of 0.1 mm. The delayed radiation was detected using three
Hammamatsu avalanche photo diodes (APD) behind one another in order to
increase detecting efficiency. The\emph{\ }delayed photons were time
integrated using the time windows given by $t_{1}=1.97~\mathrm{ns}$ and
$t_{2}=21.63~\mathrm{ns}$. For these special APDs the dead time was as small
as below $t_{1}$. Fig. \ref{expsim2D} shows the two-dimensional experimental
(a) and simulated (b) '$\omega-2\theta$' SMR\ maps in the vicinity of the
$1/2$-order\ Bragg position. In order to avoid transformation of the angles
varied in the experiment, instead of the $\theta_{\mathrm{in}}-\theta
_{\mathrm{out}}~$plane, data and simulations are displayed in the
$\omega-2\theta$ plane. Along the diagonal, the specular line was added taking
the experimental receiving slit width into account. By trial and error and
visual comparison, the best correlation function (\ref{36}) was found with the
correlation length of $\xi=1.0\ \mathrm{\mu m}$. The simulation is in rather
good agreement with the experimental data. A detailed discussion of
experimental diffuse SMR scans including external field and field-history
dependence of the domain structure will be reported elsewhere.

\section{Summary}

In summary, expressions for the diffuse scattering intensity of
grazing-incidence nuclear resonant scattering of synchrotron as well as
polarized neutron radiation have been derived in a distorted-wave
approximation. Distortion only of the incident wave was taken into account. In
a common optical formalism,\cite{Deak01} grazing-incidence (specular) x-ray,
polarized neutron, nuclear resonance reflection, and grazing-incidence diffuse
scattered intensity were calculated in terms of (geometrical) domain
correlation functions and the specular field depth profile. The formula
describes scattering by domains of rather general types with intra- and
inter-plane correlations and is not restricted to the treatment of random
lateral roughness. It was shown that, since the off-specular scatter shape is
strongly dependent on the in-plane correlation length of a single exponential
correlation function, without properly accounting for the specular field depth
profile, no conclusions can be drawn on the shape of the domain correlation
function. By prescribing the reciprocity (mirror symmetry of the
$\theta_{\mathrm{in}}-\theta_{\mathrm{out}}$ maps) the inadequacy of DIWA for
critical exit angles was eliminated. In addition, reciprocity was shown to
exacly hold for the most widely used case of SMR, i.e., when the layer
magnetizatios are parallel/antiparallel to the photon wave vector. The code
based on the presented theory is suitable for simulation of diffuse SMR maps
in a feasible calculation time, which is not yet the case for the DWBA
formulae presented in Appendix \ref{Appendix3}. Off-specular SMR and PNR
$\omega$-scans and $\theta_{\mathrm{in}}-\theta_{\mathrm{out}}$ maps of an
antiferromagnetic $\left[  \mathrm{Fe}/\mathrm{Cr}\right]  $ multilayer were
calculated and compared to each other in order to show the different features
of diffuse scattering of nuclear resonant synchrotron radiation and polarized
neutrons. As an application of the presented theory, $\omega-2\theta$ nuclear
resonant diffuse scattering maps of an $\mathrm{MgO}/\left[  ^{57}%
\mathrm{Fe}\left(  2.62\,\mathrm{nm}\right)  /\mathrm{Cr}\left(
1.28\,\mathrm{nm}\right)  \right]  _{20}$\emph{\ }multilayer\emph{\ }were
simulated and, using an exponential in-plane correlation function, a layer
magnetization correlation length of $\xi=1.0\ \mathrm{\mu m}$ was derived for
the ML demagnetized from easy axis saturation to remanence.

\appendix

\section{General solution of the coherent field equation\label{Appendix1}}

The solution of Eq. (\ref{3}) was given in
Refs.\cite{Deak01,Deak96,Spiering00} where, using the derivative field
$\Phi_{\mathrm{coh}}\left(  r_{\perp}\right)  =\left(  ik\sin\theta\right)
^{-1}\Psi_{\mathrm{coh}}^{\prime}\left(  r_{\bot}\right)  $, the second-order
differential equation regarding to $\Psi_{\mathrm{coh}}\left(  \mathbf{r}%
\right)  $, was replaced by a set of first-order differential
equations,\cite{Deak01} providing the solution
\begin{equation}
\binom{\Phi\left(  k_{\perp},r_{\perp}\right)  }{\Psi\left(  k_{\perp
},r_{\perp}\right)  }=L\left(  k_{\perp},r_{\perp}\right)  \binom{\Phi\left(
k_{\perp},0\right)  }{\Psi\left(  k_{\perp},0\right)  }, \label{A1}%
\end{equation}
where $L$ is the $4\times4$ characteristic matrix of the
system,\cite{Deak01,Deak96,Spiering00} $k_{\perp}=k\sin\theta$ is the
plane-perpendicular component of the wave number vector of the incident plane
wave, which latter dependence we drop in this appendix. Here $\Psi$ and $\Phi$
are coherent fields; the notation 'coh' has been dropped. The physical meaning
of Eq. (\ref{A1}) is that there exists a linear connection expressed by the
characteristic matrix $L$ between the fields at depth $r_{\perp}=0$ and at an
arbitrary depth $r_{\perp}$. Taking into account the boundary conditions, the
field at the top surfaces of the system $\left(  r_{\perp}=0\right)  $ is
\begin{equation}
\Psi\left(  0\right)  =\Psi^{\mathrm{in}}+R_{\mathrm{sp}}\Psi^{\mathrm{in}%
}\text{,} \label{A2}%
\end{equation}
\textit{i.e}., the sum of the incident $\Psi^{\mathrm{in}}$ and the reflected
$R_{\mathrm{sp}}\Psi^{\mathrm{in}}$ waves so that Eq. (\ref{A1}) reads
\begin{equation}
\binom{\Phi\left(  r_{\perp}\right)  }{\Psi\left(  r_{\perp}\right)
}=L\left(  r_{\perp}\right)  \binom{\Psi^{\mathrm{in}}-R_{\mathrm{sp}}%
\Psi^{\mathrm{in}}}{\Psi^{\mathrm{in}}+R_{\mathrm{sp}}\Psi^{\mathrm{in}}}
\label{A3}%
\end{equation}
where the concept of impedance tensors by Ref. \cite{Borzdov76} was used,
taking into account that the fields at $r_{\perp}=0$ are in vacuum (see Eqs.
(21) and (22) of Ref. \cite{Deak01}). Expressing the second component from Eq.
(\ref{A3}), the field at an arbitrary depth $r_{\perp}$ we have
\begin{equation}
\Psi\left(  r_{\perp}\right)  =\left\lfloor L^{\left[  21\right]  }\left(
r_{\perp}\right)  \left(  I-R_{\mathrm{sp}}\right)  +L^{\left[  22\right]
}\left(  r_{\perp}\right)  \left(  I+R_{\mathrm{sp}}\right)  \right\rfloor
\Psi^{\mathrm{in}} \label{A4}%
\end{equation}
and using the notation $T\left(  r_{\perp}\right)  =L^{\left[  21\right]
}\left(  r_{\perp}\right)  \left(  I-R_{\mathrm{sp}}\right)  +L^{\left[
22\right]  }\left(  r_{\perp}\right)  \left(  I+R_{\mathrm{sp}}\right)  $ the
solution of the three-dimensional homogeneous wave equation is
\begin{equation}
\Psi_{\mathrm{coh}}\left(  \mathbf{k,r}\right)  =T\left(  k_{\perp},r_{\perp
}\right)  \Psi^{\mathrm{in}}\exp\left(  i\mathbf{k}_{\parallel}\mathbf{r}%
_{\parallel}\right)  \text{.} \label{A5}%
\end{equation}

\section{Calculation of the $T_{l}\left(  k_{\perp},k_{\perp}^{\prime}\right)
$ Fourier integrals\label{Appendix2}}

In this appendix the analytical calculation of the integral (\ref{12}) is
given. The $4\times4$ characteristic matrix\cite{Borzdov76,Deak96} of an
arbitrary homogeneous multilayered film with layers $l=1,...,S$ is
\begin{equation}
L=L_{S}...L_{2}L_{1} \label{B1}%
\end{equation}
where%
\begin{equation}
L_{l}=\left(
\begin{array}
[c]{cc}%
\cosh\left(  kd_{l}F_{l}\right)  & \frac{1}{x}F_{l}\sinh\left(  kd_{l}%
F_{l}\right) \\
xF_{l}^{-1}\sinh\left(  kd_{l}F_{l}\right)  & \cosh\left(  kd_{l}F_{l}\right)
\end{array}
\right)  \label{B2}%
\end{equation}
is the characteristic matrix of the $l^{\mathrm{th}}$ homogeneous
layer\cite{Deak01,Deak96} with $d_{l}$ being the thickness of the
$l^{\mathrm{th}}$ layer, $x=i\sin\theta$ and the $2\times2$ matrix
$F_{l}=\sqrt{-I\sin^{2}\theta-\chi_{l}}$. We note that $L$ depends on
$k_{\perp}=k\sin\theta$, which dependence is not indicated in this Appendix.
At depth $r_{\perp}$ measured from the top of the ML the position vector
points into layer $j<S.$ The vector $r_{\perp}$ totally covers the first $j-1$
layers therefore the characteristic matrix at depth $r_{\perp}$ can be written
as%
\begin{equation}
L\left(  r_{\perp}\right)  =L_{j}\left(  r_{\perp}-D_{j-1}\right)  \cdot
L_{\left(  j-1\right)  } \label{B3}%
\end{equation}
where $D_{j-1}=%
{\textstyle\sum\limits_{l=1}^{j-1}}
d_{l}$ is the total thickness of layers up to layer $j-1$ and $L_{\left(
j-1\right)  }=L_{j-1}\cdot...\cdot L_{1}$ is the characteristic matrix of
layers $1,..,j-1.$ We note that layer $j$ is only partially covered by the
depth interval, \ which is indicated by the argument $\left(  r_{\perp
}-D_{j-1}\right)  $ in Eq. (\ref{B3}) instead of the total thickness $d_{j}$.
It is also important to note that $L\left(  r_{\perp}\right)  $ depends on the
thicknesses and susceptibilities of all covered layers and, furthermore, that
it also depends on the angle of grazing incidence $\theta$.

Using Eqs. (\ref{8}), (\ref{B2}) and (\ref{B3}) the integral (\ref{12}) can be
analytically calculated. Indeed, the two integrals
\begin{subequations}
\label{B4}%
\begin{align}
I_{j}^{+}  &  =\frac{1}{\sqrt{2\pi}}\int\limits_{Z_{j}}\mathrm{d}r_{\perp}%
\exp\left(  -ik_{\perp}^{\prime}r_{\perp}\right)  \sinh\left[  k\left(
r_{\perp}-D_{j-1}\right)  F_{j}\right] \label{B41}\\
I_{j}^{-}  &  =\frac{1}{\sqrt{2\pi}}\int\limits_{Z_{j}}\mathrm{d}r_{\perp}%
\exp\left(  -ik_{\perp}^{\prime}r_{\perp}\right)  \cosh\left[  k\left(
r_{\perp}-D_{j-1}\right)  F_{j}\right]  \label{B42}%
\end{align}
result\emph{\ }in
\end{subequations}
\begin{subequations}
\label{B5}%
\begin{align}
I_{j}^{+}  &  =\alpha_{j}+\beta_{j}\label{B51}\\
I_{j}^{-}  &  =\alpha_{j}-\beta_{j}, \label{B52}%
\end{align}
with
\end{subequations}
\begin{subequations}
\label{B6}%
\begin{align}
\alpha_{j}  &  =\exp\left(  -ik_{\perp}^{\prime}D_{j-1}\right)  \left[
\left(  K_{j}^{-}\right)  ^{-1}\exp\left(  \frac{d_{j}}{2}K_{j}^{-}\right)
\sinh\left(  \frac{d_{j}}{2}K_{j}^{-}\right)  \right] \label{B61}\\
\beta_{j}  &  =\exp\left(  -ik_{\perp}^{\prime}D_{j-1}\right)  \left[  \left(
K_{j}^{+}\right)  ^{-1}\exp\left(  -\frac{d_{j}}{2}K_{j}^{+}\right)
\sinh\left(  \frac{d_{j}}{2}K_{j}^{+}\right)  \right]  . \label{B62}%
\end{align}
Here $K_{j}^{\pm}=kF_{j}\pm ik_{\perp}^{\prime}I.$ Finally the required
expression has the form%

\end{subequations}
\begin{equation}
T_{l}\left(  k_{\perp},k_{\perp}^{\prime}\right)  =\left[  xF_{j}^{-1}%
I_{j}^{+}L_{\left(  j-1\right)  }^{\left[  11\right]  }+I_{j}^{-}L_{\left(
j-1\right)  }^{\left[  21\right]  }\right]  \left(  I-R_{\mathrm{sp}}\right)
+\left[  xF_{j}^{-1}I_{j}^{+}L_{\left(  j-1\right)  }^{\left[  12\right]
}+I_{j}^{-}L_{\left(  j-1\right)  }^{\left[  22\right]  }\right]  \left(
I+R_{\mathrm{sp}}\right)  . \label{B7}%
\end{equation}
Eq. (\ref{B7}) is physically the Fourier transform of the depth-profile
function of the coherent field and we emphasize again its dependence on
$k_{\perp}$.

\section{DWBA off-specular intensity formulae\label{Appendix3}}

As it was noted in Sections \ref{off-specular-scattering} and
\ref{off-specular-SMR}, the diffuse intensity expressions (\ref{18}) and
(\ref{25}) differ from the result of DWBA. Indeed, the DWBA transition matrix
element $\left\langle \beta\left\vert \mathbb{T}\right\vert \alpha
\right\rangle $ of the transition between the eigenstates $\left\vert
\alpha\right\rangle $ and $\left\vert \beta\right\rangle $ of the
interaction-free Hamiltonian is%
\begin{equation}
\left\langle \beta\left\vert \mathbb{T}\right\vert \alpha\right\rangle
\thickapprox\left(  \kappa_{1\beta}^{T-},V_{2}\kappa_{1\alpha}^{+}\right)  ,
\label{C1}%
\end{equation}
where $\mathbb{T}$ is the transition matrix, $V_{2}$ is the perturbing
potential, $\kappa_{1\alpha}^{+}$ and $\kappa_{1\beta}^{T-}$ are the retarded
and the advanced solutions of the unperturbed Hamiltonian and adjoint
Hamiltonian, asymptotically related to $\left\vert \alpha\right\rangle $ and
$\left\vert \beta\right\rangle $, respectively.\cite{SCHIFF} The interaction
potential $V=k^{2}\chi$ is split according to Eq. (\ref{4}) to a sum
\begin{equation}
V=V_{1}+V_{2}=k^{2}\overline{\chi}+k^{2}\left(  \chi-\overline{\chi}\right)  ,
\label{C2}%
\end{equation}
where the wave equation is exactly soluble for $V_{1}=k^{2}\overline{\chi}$
and $V_{2}=k^{2}\left(  \chi-\overline{\chi}\right)  $ is regarded as the
perturbing potential. The arguments of $\chi$ and $\overline{\chi}$ were
dropped for the sake of simplicity. According to the concept used in this
paper, the initial and final states $\left\vert \alpha\right\rangle $ and
$\left\vert \beta\right\rangle $ are plane waves with wave vectors
$\mathbf{k}$ and $\mathbf{k}^{\prime}$, respectively.

The solution of the unperturbed equation (\ref{3}), having only the $V_{1}$
potential, was given in Eq. (\ref{7}). One notes that Eq. (\ref{7}) is the
retarded solution of Eq. (\ref{3}), therefore%
\begin{equation}
\kappa_{1\alpha}^{+}=T\left(  k_{\perp},r_{\perp}\right)  \Psi^{\mathrm{in}%
}\mathrm{\exp}\left(  i\mathbf{k}_{\parallel}\mathbf{r}_{\parallel}\right)  .
\label{C3}%
\end{equation}
The advanced solution can be given similarly, taking into account that the
advanced solution is related to the emerging plane wave\emph{\ }rather than to
the incoming wave considered in Eq. (\ref{A2}). Therefore using%
\begin{equation}
\Psi\left(  0\right)  =R_{\mathrm{sp}}^{-1}\Psi^{\mathrm{out}}+\Psi
^{\mathrm{out}} \label{C4}%
\end{equation}
and $V_{1}^{\dagger}$ instead of $V_{1}$, the advanced solution can be written
as%
\begin{equation}
\kappa_{1\beta}^{T-}=\widehat{T}\left(  -k_{\perp}^{\prime},r_{\perp}\right)
\Psi^{\mathrm{out}}\mathrm{\exp}\left(  i\mathbf{k}_{\parallel}^{\prime
}\mathbf{r}_{\parallel}\right)  , \label{C5}%
\end{equation}
where $\widehat{T}$ means that in Eq. (\ref{A5}) the specular reflectivities
are inverted and the adjoint susceptibilities are used. Applying Eqs.
(\ref{C1}), (\ref{C3}), (\ref{C5}), (\ref{20}), (\ref{21}), (\ref{23}) and
(\ref{27}) the diffuse intensity is%
\begin{equation}
I_{\mathrm{off}}\left(  E\right)  =k^{4}\sum_{ll^{\prime}\mu\mu^{\prime}%
}C_{l^{\prime}l}^{\mu^{\prime}\mu}\left(  \mathbf{K}_{\parallel}\right)
\operatorname{Tr}\left[  \Gamma_{l^{\prime}}^{\mu^{\prime}}\left(  E\right)
^{\dagger}\Gamma_{l}^{\mu}\left(  E\right)  \rho\right]  \label{C6}%
\end{equation}
similarly to Eq. (\ref{25}), however, the definition of $\Gamma_{l}^{\mu}$
being
\begin{equation}
\Gamma_{l}^{\mu}\left(  k_{\perp},k_{\perp}^{\prime},E\right)  =\int
\limits_{Z_{l}}\mathrm{d}r_{\perp}\widehat{T}\left(  -k_{\perp}^{\prime
},r_{\perp}\right)  ^{\dagger}\chi_{l}^{\mu}\left(  E\right)  T\left(
k_{\perp},r_{\perp}\right)  . \label{C7}%
\end{equation}
different from that given in Eq. (\ref{26}). The integral (\ref{C7}) can be
analytically calculated following the method presented in Appendix
\ref{Appendix2}. Comparing Eq. (\ref{C7}) with Eqs. (\ref{12}) and (\ref{26})
one can realize that the present result of Eq. (\ref{25}) can be also obtained
from the DWBA by taking
\begin{equation}
\widehat{T}\left(  -k_{\perp}^{\prime},r_{\perp}\right)  ^{\dagger}\approx
I\exp\left(  -ik_{\perp}^{\prime}r_{\perp}\right)  , \label{C8}%
\end{equation}
which is valid only for exit angles ( $\theta^{\prime}$) above the critical
angle. This statement is consistent with ignoring the scattering of the
diffuse field as assumed in section \ref{off-specular-scattering}. We also
note that approximation
\begin{equation}
T\left(  k_{\perp},r_{\perp}\right)  \approx I\exp\left(  ik_{\perp}r_{\perp
}\right)  \label{C9}%
\end{equation}
together with (\ref{C8}) are identical to the conventional 1$^{\mathrm{st}}$
order Born approximation.

\begin{acknowledgments}
This work was partly supported by the Hungarian Scientific Research Fund
(OTKA) and National Office for Research and Technology of Hungary under
Contract numbers T047094 and NAP-Veneus'05 as well as by the European
Community under the Specific Targeted Research Project Contract No.
NMP4-CT-2003-001516 (DYNASYNC). The authors gratefully acknowledge the beam
time supplied free of charge by the Japan Synchrotron Radiation Institute
(JASRI) for experiment No: 2002B239-ND3-np. Our gratitude goes to \textit{A.Q.
Baron} (SPring-8, JASRI) for his kind supply of the fast Hammamatsu APD
detectors, \textit{J. Dekoster} (IKS\ Leuven) for preparing the multilayer
sample and to \textit{D. G. Merkel} (KFKI\ RMKI Budapest) for his assistance
in data processing. One of the authors (LD) gratefully acknowledges the
financial support by the Deutscher Akademischer Austauschdienst (DAAD).
\end{acknowledgments}

\newpage

\begin{thebibliography}{99}                                                                                               %


\bibitem {Stoev1999}K. N. Stoev, K. Sakurai, Spectrochimica Acta, Part B 54
(1999) 41.

\bibitem {Lax51}M. Lax, Rev. Mod. Phys. 23 (1951) 287.

\bibitem {Zhou95}X.-L. Zhou, S.-H. Chen, Phys. Rep. 257 (1995) 223.

\bibitem {Daillant}J. Daillant, A. Gibaud (eds.) "X-Ray and Neutron
Reflectivity: Principles and Applications" Lecture Notes in Physics m 58.
Springer-Verlag, New York, 1999.

\bibitem {Felcher93}G.P. Felcher, Physica B 192 (1993) 137.

\bibitem {Majkrzak91}C.F. Majkrzak, Physica B 173 (1991) 75.

\bibitem {Sinha91}S.K. Sinha, Physica B 173 (1991) 25.

\bibitem {Hannon88}J.P. Hannon, G.T. Trammell, M. Blume, and Doon Gibbs, Phys.
Rev. Lett. 61 (1988) 1245.

\bibitem {Whan94}D.B. Mac Whan, J. Synchrotron Radiat. 1 (1994) 83., and
references therein.

\bibitem {Hase00}T.P.A. Hase, I. Pape, B.K. Tanner, H. D\"{u}rr, E. Dudzik, G.
van der Laan, C.H. Marrows and B.J. Hickey, Phys. Rev. B 61 (2000) R3792.

\bibitem {Lauter00}V. Lauter-Pasyuk, H.J. Lauter, B. Toperverg, O. Nikonov, E.
Kravtsov, M.A. Milyaev, L. Romashev and V. Ustinov, Physica B 283 (2000) 194.

\bibitem {Langridge00}S. Langridge, J. Schmalian, C.H. Marrows, D.T. Dekadjevi
and B.J. Hickey, Phys. Rev. Lett. 85 (2000) 4964.

\bibitem {Nagy99}D.L. Nagy, L. Botty\'{a}n, L. De\'{a}k, E. Szil\'{a}gyi, H.
Spiering, J. Dekoster and G. Langouche, Hyp. Int. 126 (2000) 353.

\bibitem {Deak02}L. De\'{a}k, L. Botty\'{a}n, M. Major, D.L. Nagy, H.
Spiering, E. Szil\'{a}gyi and F. Tanczik\'{o}, Hyp. Int. 144/145 (2002) 45.

\bibitem {Toellner95}T.S. Toellner, W. Sturhahn, R. R\"{o}hlsberger, E.E. Alp,
C.H. Sowers and E.E. Fullerton, Phys. Rev. Lett. 74 (1995) 3475.

\bibitem {Chumakov99}A.I. Chumakov, L. Niesen, D.L. Nagy and E.E. Alp, Hyp.
Int. 123/124 (1999) 427.B.

\bibitem {Rohlsberger03}R. R\"{o}hlsberger, J. Bansmann, V. Senz, K.L. Jonas,
A. Bettac, K.H. Meiwes-Broer, O. Leupold, Phys. Rev. B. 67 (2003) 245412.

\bibitem {Mandel-Wolf}L. Mandel and E. Wolf, "Optical coherence and quantum
optics" p 155, (Cambridge University Press, 1995)

\bibitem {Toperverg01a}B.T. Toperverg, Physica B 297 (2001) 160.

\bibitem {Nagy02a}D.L. Nagy, L. Botty\'{a}n, B. Croonenborghs, L. De\'{a}k, B.
Degroote, J. Dekoster, H.J. Lauter, V. Lauter-Pasyuk, O. Leupold, M. Major, J.
Meersschaut, O. Nikonov, A. Petrenko, R. R\"{u}ffer, H. Spiering and E.
Szil\'{a}gyi, Phys. Rev. Lett. 88 (2002) 157202.

\bibitem {Deak01}L. De\'{a}k, L. Botty\'{a}n, D.L. Nagy, and H. Spiering,
Physica B 297 (2001) 113.

\bibitem {Afanasjev65}A.M. Afanas'ev and Yu. Kagan, Sov. Phys. ---JETP 21
(1965) 215.

\bibitem {Hannon69}J.P. Hannon and G.T. Trammell, Phys. Rev. 186 (1969) 306.

\bibitem {Hannon85}J.P. Hannon, G.T. Trammell, M. Mueller, E. Gerdau, R.
Ruffer, and H. Winkler, Phys. Rev. B 32 (1985) 6363.

\bibitem {Irkajev93}S.M. Irkaev, M.A. Andreeva, V.G. Semenov, G.N. Belozerskii
and O.V. Grishin, Nucl. Instrum. Methods B 74 (1993) 545.

\bibitem {Irkajev93b}S.M. Irkaev, M.A. Andreeva, V.G. Semenov, G.N.
Beloserskii, and O.V. Grishin, Nucl. Instrum. Methods B 74 (1993) 554.

\bibitem {Deak96}L. De\'{a}k, L. Botty\'{a}n, D.L. Nagy and H. Spiering, Phys.
Rev. B. 53 (1996) 6158.

\bibitem {Rohlsberger99a}R. R\"{o}hlsberger, Hyp. Int. 123/124 (1999) 301.

\bibitem {Rohlsberger99b}R. R\"{o}hlsberger, Hyp. Int. 123/124 (1999) 455.

\bibitem {Sinha88}S.K. Sinha, E.B. Sirota, S. Garoff, H.B. Stanley, Phys. Rev
B 38 (1988) 2297.

\bibitem {Toperverg01b}B.T. Toperverg, O. Nikonov, V. Lauter-Pasyuk, H.J.
Lauter, Physica B 297 (2001) 169.

\bibitem {Lauter94}V. Lauter-Pasyuk at al., J. Magn. Magn. Mater. 226-230
(2001) 1694.

\bibitem {Ruhm99}A. R\"{u}hm, B.P. Toperverg, H. Dosch, Phys. Rev. B. 60
(1999) 16073.

\bibitem {Vineyard82}G. H. Vineyard, Phys. Rev. B. 26 (1982) 4146.

\bibitem {Dosch}H. Dosch, \textquotedblleft Critical Phenomena at Surfaces and
Interfaces\textquotedblright\ (Springer-

Verlag, New York 1992).

\bibitem {Ljungsdahl}G.Ljungdahl~and~S.W.Lovesey,~Physica~Scripta~53~(1996)~734.

\bibitem {Felcher87}G.P. Felcher, R.O. Hilleke, R.K. Crawford, J. Haumann, R.
Kleb, and G. Ostrowski, Rev. Sci. Instrumm 58 (1987) 609.

\bibitem {Paratt54}L.G. Parratt, Phys. Rev. 95 (1954) 359.

\bibitem {Majkrzak89}C.F. Majkrzak, Physica B 156$\backslash$\&157 (1989) 619.

\bibitem {Nickel2001}B. Nickel, A. R\"{u}hm, W. Donner, J. Major, H. Dosch, A.
Schreyer, H. Zabel, and H. Hublot, Rev. Sci. Instrum 72 (2001) 163.

\bibitem {Hannon85b}J.P. Hannon, N.V. Hung, G.T. Trammell, E. Gerdau, M.
Mueller, R. R\"{u}ffer, and H. Winkler, Phys. Rev. B 32 (1985) 5068.

\bibitem {Blume68}M. Blume and O.C. Kistner, Phys. Rev. 171 (1968) 417.

\bibitem {Trammell79}G.T Trammell and J.P. Hannon, Phys. Rev B 18 (1978) 165.

\bibitem {SCHIFF}Leonard I. Schiff, "Quantum mechanics" p 327 (McGraw-Hill, 1955).

\bibitem {BornWolf}M. Born and E. Wolf, "Princeples of optics", (Cambridge
University Press, 7th edition, 1999)

\bibitem {Potton2004}R.J. Potton, Rep. Prog. Phys \textbf{67} (2004) 717.

\bibitem {Henke}L. Henke, Atomic data and nuclear data tables 54, 1993, 181;
see also http://henke.lbl.gov/optical\_constants/getdb2.html

\bibitem {Neutron News}Neutron News, Vol. 3, No. 3, 1992, pp. 29-37, http://www.ncnr.nist.gov/resources/n-lengths/

\bibitem {Spiering00}H. Spiering, L. De\'{a}k, and L. Botty\'{a}n, Hyp. Int.
125 (2000) 197.

\bibitem {EFFI}The computer program EFFI is available from ftp://nucssp.rmki.kfki.hu/pub/effi

\bibitem {Paul}A. Paul, E. Kentzinger, U. R\"{u}cker, D. E. B\"{u}rgler, and
Thomas Br\"{u}ckel, Phys. Rev. B \textbf{73}, (2006) 94441.

\bibitem {Baron96}A. Q. R. Baron, A. I. Chumakov, H. F. Gr\"{u}nsteudel, H.
Gr\"{u}nsteudel, L. Niesen, and R. R\"{u}ffer, PRL 77 (1996) 4808.

\bibitem {Sinha1998}S.K. Sinha, M. Tolan, A. Gibaud, Phys. Rev. B \textbf{57}
(1998) 2740.

\bibitem {Lovesely-II}S. W. Lovesey, "Theory of Neutron Scattering from
Condensed Matter

Volume II: Polarization Effects and Magnetic Scattering" Clarendon Press, 1986.

\bibitem {Kiessig31}H. Kiessig, Annalen der Physik (1931) 769.

\bibitem {Deak94}L.~De\'{a}k, L. Botty\'{a}n, D.L.~Nagy, Hyp. Int. 92 (1994) 1083.

\bibitem {Baron94a}A.Q.R. Baron, J. Arthur, S.L. Ruby, A.I. Chumakov, G.V.
Smirnov, G.S. Brown, Phys. Rev. B \textbf{50,} (1994) 10354.

\bibitem {Machine-time}The calculation time for the diffuse SMR map
$100\times100$ points in Fig. \ref{twod-smr} on a 64-bit PC with 1024 Mb RAM
and AMD Athlon 3000+ processor running a single user process under SuSe Linux
10.0 was 12 hours. The estimated DWBA calculation time is exactly one week.

\bibitem {Huffman1970}A.H. Huffman, Phys. Rev. D \textbf{1}, (1970) 890.

\bibitem {reciprocity}The proof of the corresponding reciprocity
\textit{theorem} is a subject of a number of theoretical
work.\cite{BornWolf,SCHIFF,Potton2004,Saxon1955,Hillion1978,Carminati2000}%
\emph{\ }Rigorous proof was given for the cases of real\cite{SCHIFF,Saxon1955}
\emph{\ }(like in PNR) and for complex and short range,
polarization-independent\cite{BornWolf,SCHIFF,Carminati2000} scattering
potentials.\cite{Huffman1970} Whether or not reciprocity actually holds for
the case of nuclear resonant photons, where the susceptiblilty and,
consequently, $k^{2}\chi$, the scattering potential is complex and
polarization dependent, is beyond the scope of the present discussion.

\bibitem {Chernov2005}V.A. Chernov, V.I. Kondratiev, N.V. Kovalenko, S.V.
Mytnichenko, K.V. Zolotarev, Physica B \textbf{357}, (2005) 232.

\bibitem {Mytnichenko2005}S.V. Mytnichenko, Physica B \textbf{355}, (2005) 244.

\bibitem {Chernov2000}V.A. Chernov, E.D. Chkhalo, N.V. Kovalenko, S.V.
Mytnichenko, Nuclear Instruments and Methods in Physics Research A
\textbf{448}, (2000) 276.

\bibitem {Chernov2003}V.A. Chernov, N.V. Kovalenko, S.V. Mytnichenko and A.I.
Toropov, Act. Crist. A 59, (2003) 551.

\bibitem {BottyanBSF1}L. Botty\'{a}n, L. De\'{a}k, J. Dekoster, E. Kunnen, G.
Langouche, J. Meersschaut, M. Major, D.L. Nagy, H.D. R\"{u}ter, E.
Szil\'{a}gyi, K. Temst, J. Magn. Magn. Mater. \textbf{240}, (2002) 514.

\bibitem {Nagy02b}D.L. Nagy, L. Botty\'{a}n, L. De\'{a}k, B. Degroote, O.
Leupold, M. Major, J. Meersschaut, R. R\"{u}ffer, E. Szil\'{a}gyi, J. Swerts,
K. Temst, Phys. Stat. Sol. (a) 189 (2002) 591.

\bibitem {Tancziko2004}F. Tanczik\'{o}, L. De\'{a}k, D.L. Nagy, L.
Botty\'{a}n, Nucl. Instr. Meth. Phys. Res. B \textbf{226}, (2004) 461.

\bibitem {Yoda2001}Y. Yoda, M. Yabashi, K. Izumi, X.W. Zhang, S. Kishimoto, S.
Kitao, M. Seto, T. Mitsui, T. Harami, Y. Imai, and S. Kikuta, Nucl. Instrum.
Methods Phys. Res. A \textbf{467}, (2001) 715.

\bibitem {Borzdov76}G.M. Borzdov, L.M. Barskovskii, and V.I. Lavrukovich, Zh.
Prikl. Spektrosk. 25 (1976) 526.

\bibitem {Saxon1955}David S. Saxon, Phys. Rev. \textbf{100} (1955) 1771.

\bibitem {Hillion1978}P. Hillion, J. Optics \textbf{9} (1978) 173.

\bibitem {Carminati2000}R. Carminati, J.J. S\'{a}enz, J.-J. Greffet, and M.
Nieto-Vesperinas, Phys. Rev. \textbf{A} 62 (2000) 012712
\end{thebibliography}
\end{document}